\documentclass[11pt,a4paper]{article}
\pdfoutput=1
\usepackage{jheppub}

\usepackage{amsmath}
\usepackage{verbatim}
\usepackage{graphicx}
\usepackage{mathrsfs}
\usepackage{appendix}
\usepackage{caption}
\usepackage{float}
\usepackage{subfig}
\usepackage{mathtools}
\usepackage{tikz-feynman}
\usepackage{setspace}

\newcommand{\g}{\gamma}
\newcommand{\gb}{\hat{\gamma}}
\newcommand{\gc}{\tilde{\gamma}}
\newcommand{\sg}{\sigma}
\newcommand{\hs}{\hat{\Sigma}}
\newcommand{\ts}{\tilde{\Sigma}}

\newcommand{\be}[1]{ \begin{equation}\label{#1} }
\newcommand{\ee}{\end{equation}}
\newcommand{\bea}[1]{\begin{eqnarray}\label{#1} }
\newcommand{\eea}{\end{eqnarray}}
\newcommand{\bes}{\begin{subequations}}
\newcommand{\ees}{\end{subequations}}

\newcommand{\p}{\partial}
\renewcommand{\a}{\alpha}

\newcommand{\s}{\sigma}

\newcommand{\refb}[1]{(\ref{#1})}

\newcommand{\non}{\nonumber}

\newcommand{\rw}{\rightarrow}

\title{Magic Fermions: Carroll and Flat Bands}

\author{}

\author[a,b]{Arjun Bagchi,} \author[c]{Aritra Banerjee,} \author[d]{Rudranil Basu,} \author[a]{Minhajul Islam,} \author[a]{Saikat Mondal} \author{\\} 

\affiliation[a]{Indian Institute of Technology Kanpur, Kalyanpur, Kanpur 208016. INDIA. \\}

\affiliation[b]{Centre de Physique Theorique, Ecole Polytechnique de Paris, 91128 Palaiseau Cedex, France.\\}

\affiliation[c]{Okinawa Institute of Science and Technology (OIST),
1919-1 Tancha, Onna-son, Okinawa 904-0495, JAPAN\\}

\affiliation[d]{Department of Physics, BITS-Pilani, K K Birla Goa Campus, Zuarinagar, Goa-403726. INDIA.\\}

\emailAdd{abagchi@iitk.ac.in, aritra.banerjee@oist.jp, rudranilb@goa.bits-pilani.ac.in, saikatmd@iitk.ac.in, minhajul@iitk.ac.in}

\abstract{The Carroll algebra is constructed as the $c\to0$ limit of the Poincare algebra and is associated to symmetries on generic null surfaces. In this paper, we begin investigations of Carrollian fermions or fermions defined on generic null surfaces. Due to the availability of two different (degenerate) metrics on Carroll spacetimes, there is the possibility of two different versions of Carroll Clifford algebras. We consider both possibilities and construct explicit representations of Carrollian gamma matrices and show how to build higher spacetime dimensional representations out of lower ones. Actions for Carroll fermions are constructed with these gamma matrices and the properties of these actions are investigated. 

\medskip

We show that in condensed matter systems where the dispersion relation becomes trivial i.e. the energy is not dependent on momentum and bands flatten out, Carroll symmetry generically appears. We give explicit examples of this including that of twisted bi-layer graphene, where superconductivity appears at so called magic angles and connect this to Carroll fermions.}

\preprint{}

\begin{document}
	
\maketitle
	
\newpage

\section{Introduction}

{\it{``Everyone wants a magical solution to their problem, and everyone refuses to believe in magic."}}

\medskip

The physics of flat bands in condensed matter has attracted a lot of recent interest with wide ranging applications in exotic systems like fractional quantum Hall systems, spin liquids and twisted bi-layer graphene where ``magic" superconductivity appears at certain specific angles of twist. In this paper, we will propose a generic solution to flat bands that has a touch of magic, --- fermions with Carrollian symmetry.

\subsection*{The rise and rise of Carroll}
Non-Lorentzian physics or the study of physics beyond relativistic invariance and without Lorentz or Poincare symmetry and the geometry underlying these Non-Lorentzian structures have become very fashionable of late. This is primarily due to the discovery of uses of Non-Lorentzian structures in various branches of physics including condensed matter physics, classical and quantum gravity. In this work, we will be interested in Carrollian theories, where the Carroll group replaces the Poincare group as the symmetry group of interest. 

\medskip

Starting from Poincare group, one can obtain Carroll group as a contraction by taking the speed of light going to zero limit. This was first introduced by Levy-Leblond \cite{LevyLeblond}  and Sen Gupta \cite{NDS} as a ``degenerate cousin of the Poincare group". 
Numerous counter-intuitive things occur in this limit. The light-cone closes completely, the space-time metric becomes degenerate and new non-Lorentzian structures emerge on the spacetime manifold \cite{Henneaux:1979vn}. These Carrollian manifolds can be understood mathematically as fibre-bundles where there are two distinct metrics on the base and the fibre instead of a single non-degenerate metric for the entire spacetime. Interestingly, any null hypersurface is a Carroll manifold. Two of the most important examples of null hypersurfaces are null boundary of asymptotically flat spacetimes and event horizons of generic black holes. Thus Carrollian and conformal Carrollian structures are inexorably linked with scatterings in flat spacetimes and black hole physics. 

\medskip

Conformal Carrollian symmetry has been discovered to be isomorphic to Bondi-Metzner-Sachs (BMS) symmetry in generic dimensions \cite{Duval:2014uva}, following closely related observations in two dimensions \cite{Bagchi:2010zz}. BMS groups emerge as the asymptotic symmetry groups for asymptotically flat spacetimes at null infinity \cite{Bondi:1962px,Sachs:1962zza}. These groups are infinite-dimensional in dimensions three \cite{Barnich:2006av} and four {\footnote{In higher dimensions, depending on boundary conditions the BMS can be truncated to the Poincare algebra or can still remain infinite dimensional \cite{Kapec:2015vwa}.}}. A considerable body of research has been built up relating asymptotic symmetries, soft theorems and memory effect following the lead of Strominger \cite{Strominger:2013jfa}. The formulation of holography for asymptotically flat spacetimes has followed two routes, Celestial and Carrollian holography. In the Celestial formulation, it has been proposed that a 2d conformal field theory living on the celestial sphere can compute scattering amplitudes in 4d asymptotically flat spacetimes. This has led to novel results about asymptotic symmetries and soft theorems. See \cite{Strominger:2017zoo, Pasterski:2021rjz,Raclariu:2021zjz} for recent reviews on Celestial holography and references therein. Carroll holography on the other hand postulates codimension one holography between flat spacetimes and a Carrollian CFT living on the null boundary of flatspace \cite{Bagchi:2010zz, Bagchi:2012cy, Bagchi:2016bcd}. This has seen quite a bit of success, particularly in lower dimensions. See \cite{Bagchi:2012xr,Barnich:2012xq,Bagchi:2013qva,Bagchi:2015wna,Bagchi:2014iea,Jiang:2017ecm,Barnich:2012aw,  Duval:2014lpa, Hartong:2015usd} for a non-exhaustive list of important contributions in this regard. The connection between 4d scattering and 3d Carrollian CFTs has been recently proposed in \cite{Bagchi:2022emh} drawing inspiration from the Celestial approach \footnote{See also \cite{Donnay:2022aba} for further connections between the Celestial and Carrollian formulations of flat holography.}.

\medskip

We have mentioned above that Carrollian structures emerge naturally on all null surfaces. The event horizons of generic black holes form an important class of these null surfaces and defining theories on these null surfaces naturally gives rise of Carrollian (conformal) field theories. The emergence of Carrollian and BMS symmetries (as well as ``BMS-like" symmetries) on black hole even horizons have been addressed in \cite{Donnay:2019jiz,Hawking:2016msc,Donnay:2015abr,Afshar:2016wfy,Penna:2017bdn,Carlip:2017xne}. If one wishes to define quantum field theories living on black hole event horizons, it is but natural that these theories would imbibe the Carroll structures that appear geometrically on these degenerate backgrounds. The investigation of Carrollian field theories become important in this context as well.

\medskip

Carroll symmetry also appears in the tensionless limit of string theory. In this limit the worldsheet metric becomes degenerate and BMS$_3$ replaces the worldsheet conformal symmetry \cite{Bagchi:2013bga,Bagchi:2015nca,Bagchi:2019cay,Bagchi:2020fpr,Bagchi:2020ats}.
It has been shown in \cite{Bagchi:2021ban} that strings probing black hole geometries become effectively tensionless near the horizon of the black hole. The background Carroll structure of the blackhole event horizon induces a Carroll structure on the worldsheet of the string. Rather intriguingly, in a recent paper \cite{Bagchi:2022iqb}, it has been shown that null strings wrapping the horizon of a BTZ black hole in asymptotically AdS spacetimes can give an explicit counting of its microstates, reproducing not only the leading Bekenstein-Hawking entropy but amazingly even its logarithmic corrections correctly.   

\medskip

A brief word on the other applications. A substantial amount of research has been done regarding Carroll aspects of gravity \cite{Bergshoeff:2017btm,Duval:2017els,Hartong:2015xda,Morand:2018tke,Ciambelli:2018ojf,Henneaux:2021yzg}. Carroll fluids have been addressed in \cite{deBoer:2017ing,Ciambelli:2018wre,Ciambelli:2018xat,Ciambelli:2020eba,Petkou:2022bmz,Freidel:2022bai,Freidel:2022vjq}. New links have emerged between Carroll symmetry and cosmology and dark energy \cite{deBoer:2021jej}, as well as connections to condensed matter physics through the theory of fractons \cite{Bidussi:2021nmp}. 

\subsection*{In this paper: Carroll Fermions and Flat Bands}

There has been many recent attempts at understanding various aspects of field theories having Carrollian and Conformal Carrollian symmetry. Initially the effort was to understand these as the systematic $c\to0$ limit of relativistic field theories \cite{Bagchi:2019xfx} and the Carrollian version of scalars, electrodynamics and Yang-Mills theories were constructed and studied. Later, there has been the effort in constructing actions for these theories \cite{Bergshoeff:2017btm, Bagchi:2019clu} and studying theories from these actions. Scalars \cite{Gupta:2020dtl, deBoer:2021jej, Bagchi:2022eav, Baiguera:2022lsw, Rivera-Betancour:2022lkc}, gauge fields, and gravity \cite{Bergshoeff:2017btm,Duval:2017els,Hartong:2015xda,Morand:2018tke,Ciambelli:2018ojf,Henneaux:2021yzg} have received much attention. Rather surprisingly, fermions have been left out of the fun {\footnote{See however \cite{Bergshoeff:2017btm} for an earlier attempt (we will comment briefly about this in Section \eqref{sec4}) and \cite{Yu:2022bcp,Hao:2022xhq} for very recent explorations in $d=2$.}}. In this paper we aim to rectify this and build towards an understanding of Carroll fermions in general dimension $d$. A companion paper \cite{2dcarroll} contains details of Carroll fermions specifically for $d=2$.  
\medskip

In condensed matter physics, one of the very important notions is that of the dispersion relation which captures the dependence of energy with spatial momentum in a system. The very simple non-relativistic dispersion relation $E=\frac{p^2}{2m}$ captures complex phenomena like Bose-Einstein condensation. Of late, there has been a great deal of interest in theories having trivial dispersion relations, where energies don't depend on momentum at all. These so called flat band structures are at the heart of many exotic phenomena including fractional quantum Hall effect, and magic superconductivity in twisted bi-layer graphene. We will show in this paper that the appearance of flat bands is directly connected to the appearance of Carrollian symmetry in the systems under consideration. As these systems are usually fermionic systems, the Carrollian theories that are connected to these are the Carrollian fermions we build in the first part of this paper.  

\medskip

The rest of the paper is organised as follows. We give a short review of Carroll symmetry from an algebraic and a geometric perspective in Section \eqref{sec2}. Section \eqref{sec3} is our first encounter with Carroll fermions. Here we define the Carroll Clifford algebra and find two distinct types of algebras and discuss their properties and representations. In Section \eqref{sec4}, we build actions for the two different fermions and discuss structures and symmetries related to the actions. In Section \eqref{sec5}, we move on to show how Carroll symmetry emerges in flat band materials in condensed matter systems and follow that with explicit examples in Section \eqref{sec6}. We conclude with some remarks in Section \eqref{sec7}. An appendix has details of Carrollian and conformal Carrollian isometries. 
	
\newpage	
	
\section{Carroll Symmetry}\label{sec2}
In this section we will give a quick recap of Carroll symmetry. The Carroll group is be obtained by taking the speed of light going to zero limit of the relativistic Poincare symmetry. This is the algebraic approach we go on to describe below. There is also a geometric way of describing the Carroll group in terms of degenerate metrics and Carroll manifolds. We will elaborate on this as well. 
	\subsection{Contraction, transformations and algebra}
	
	Carroll group arises from an Inonu-Wigner contraction from the Poincare group. This is performed by taking the limit $c\rw0$ of the Poincare group. This is equivalent to the scaling
	\begin{equation}
		x^i\rw  x^i ,\quad t\rw \epsilon t, \quad \epsilon\rw 0
	\end{equation}
	where $i = 1,...,(d-1)$. Under this limit the Carroll generators are defined by re-scaling the parent Poincare generators in the following way 
\be{}
H \equiv \epsilon P_0^{\text{P}}, \, P_i \equiv P_i^{\text{P}}, \, C_i \equiv \epsilon J_{0i}^{\text{L}}, \, J_{ij}\equiv J_{ij}^{\text{P}},
\ee
where the superscript ${\text{P}}$ refers to the Poincare generators. The explicit form of the Carroll generators are:
\begin{equation}
H=\p_{t} ,\quad P_i=\p_{i} ,\quad C_i=x_i\p_{t} ,\quad J_{ij}=x_i\p_{j}-x_j\p_{i}.
\end{equation}
Here $H,P_i,C_i,J_{ij}$ are time translation, spatial translations, Carroll boosts and $(d-1)$ spatial rotations respectively. These generators generate the Carroll transformation on space-time coordinate
\be{}		
t^{'}=t+a -\vec{v}.\vec{x}, \quad \vec{x}^{'}=\boldsymbol{R}\vec{x}+\vec{b}
\ee
where the parameters of the group $(a, \vec{b}, \vec{v}, \boldsymbol{R})$ describes time-translation, space-translation, Carroll boosts and $SO(d-1)$ rotation respectively. The Lie algebra of the Carroll group is given by following non-zero commutation relations
\begin{eqnarray}\label{carral}
[J_{ij},J_{kl}]=4\delta_{[i[k}J_{l]j]}, \, [J_{ij},P_{k}]=2\delta_{k[j}P_{i]}, \, [J_{ij},C_{k}]=2\delta_{k[j}C_{i]}, \, [C_i,P_j]=-\delta_{ij}H. 
\end{eqnarray}
Crucially, the commutation relation between the Carroll boosts becomes $[C_i, C_j]=0$, reflecting the non-Lorentzian nature of the algebra. When we generalise to conformal Carroll symmetry we have additional generators, Dilatation $(D)$, temporal $(K_0)$ and spatial $(K_i)$ special Conformal Carroll transformation:
\begin{equation}
D=t\p_{t} + x_i\p_{i} ,\quad K_0 = x_ix_i\p_{t} ,\quad K_i = 2x_i(t\p_{t} + x_j\p_{j}) - x_jx_j\p_{i}.
\end{equation}
The additional non-vanishing commutation relations are 
\begin{eqnarray}
&&[D,P_i] = -P_i, \, [D,H] = -H \, [D,K_i] = K_i, \, [D,K_0] = K_0, \nonumber\\
&&[K_0,P_i] = -2C_i \, [K_i,H]=-2C_i, \, [K_i,P_j] = -2\delta_{ij}D-2J_{ij}. 
\end{eqnarray}
This is the finite Conformal Carrollian Algebra (f-CCA). The algebra can be obtained by the $c\to0$ limit of the relativistic Conformal algebra.  
It is important to note that this finite algebra can be rewritten in a suggestive form and be given an infinite dimensional lift {\em for all spacetime dimensions} \cite{Bagchi:2016bcd}. This infinite dimensional lift makes connection with the important fact that CCA is isomorphic to the BMS algebra in one higher dimension
\be{}
\mathfrak{CCarr}_d = \mathfrak{bms}_{d+1}.
\ee
The BMS algebra is the symmetry of the null boundary of asymptotically flat spacetime in $(d+1)$ dimensions. The topology of null boundary is $R_u\otimes S^d$, where $R_u$ is a null line.  

The details of the infinite lift of the CCA are dimension dependent. The infinite extension in $d=2$ gives the BMS$_3$ algebra:
\begin{align}
& [L_n, L_m] = (n-m)L_{n+m} + \frac{c_L}{12}(n^3-n)\delta_{n+m,0}, \cr
& [L_n, M_m] = (n-m)M_{n+m} + \frac{c_M}{12}(n^3-n)\delta_{n+m,0}, \cr
& [M_n, M_m]=0
\end{align}
Here the supertranslations $M_n$ are the generators of angle-dependent translations of the null direction $u$ and $L_n$, the superrotations, generate the diffeomorphism of the circle at infinity. There is a similar enhancement possible for $d=3$ which gives rise to the BMS$_4$ algebra. Even for $d\geq4$ it is possible to give an infinite extension for the supertranslations \cite{Bagchi:2016bcd}. For $d=4$ this is 
\begin{equation}
	M_f = f(x,y,z)\p_{t},
\end{equation}
for any arbitrary tensor field $f$, transforming under $SO(3)$. These close to form an infinite Abelian ideal along with the (finite) conformal generators on the sphere. It is possible to have an extension to the full diffeomorphism group on the sphere with other boundary conditions, but we will not be interested in this and other generalisations of the BMS group in this paper. 

\subsection{Carroll Geometry}

Let us now give a more geometric picture of Carroll symmetry and review the non-Lorentzian structures that arise in this context. 

\medskip

We begin by considering the Carroll limit of ordinary Minkowski spacetime to get a first idea of the degenrate structures we would encounter. Consider the $d$-dimensional Minkowski space with line element:
	\begin{equation}
		ds^2 =-c^2dt^2+(dx^i)^2
	\end{equation}
	Then the covariant metric and its contravariant inverse are: 
	\begin{eqnarray}
		\eta_{\mu\nu} = 
		\begin{pmatrix}
			-c^2  &0 \\
			0  &I_{d-1}
		\end{pmatrix}
		\qquad	\eta^{\mu\nu} =
		\begin{pmatrix}
			-1/c^2 & 0 \\
			0 & I_{d-1} 
		\end{pmatrix} 
	\end{eqnarray}
	Now taking the Carroll limit $(c\rw0)$ we get a degenerate covariant spatial metric $\tilde{h}_{\mu\nu}$ with one zero eigenvalue and a degenerate contravariant temporal metric $\Theta^{\mu\nu}$ with one non-zero eigenvalue: 
\be{}
\eta_{\mu\nu} \rightarrow \tilde{h}_{\mu\nu} = \begin{pmatrix}
			0 & 0 \\
			0 & \quad I_{d-1}
		\end{pmatrix}, 
		\qquad
		-c^2 \eta^{\mu\nu} \rightarrow \Theta^{\mu\nu} = 	\begin{pmatrix}
			1 & 0 \\
			0 & \quad 0_{d-1}
		\end{pmatrix}   \ee		
These two are the invariant tensors for the Carroll group. As $\Theta^{\mu\nu}$ is basically 1$\times$1 matrix, we can define vector $\theta^{\mu}$ such that 
\be{}
\Theta^{\mu\nu}=\theta^{\mu}\theta^{\nu}
\ee and degeneracy implies 
\be{} \tilde{h}_{\mu\nu}\theta^{\nu}=0.
\ee
	
\medskip
	
Having had a brief encounter with the degeneracy we will encounter in the limit, let us now see how we can formalise this structure from an instrinsically Carrollian point of view \cite{Henneaux:1979vn}. To this end, we now review Carroll geometry following \cite{Duval:2014uoa}. A Carroll structure is a quadruplet $(\mathcal{C},\tilde{h},\theta, \nabla)$, where
\begin{itemize}
\item $\mathcal{C}$ is a $d$ dimensional manifold, on which one can choose a coordinate chart $(t,x^{i})$. 
\item $\tilde{h}$ is a covariant, symmetric, positive,  tensor field of rank $d-1$ and of signature $(0,\underbrace{+1,\ldots,+1}_{d-1})$.
\item $\theta$ is a non-vanishing vector field which generates the kernel of $\tilde{h}$ . 
\item $\nabla$ is a symmetric affine connection that parallel transports both $\tilde{h}_{\mu\nu}$ and $\theta^{\nu}$. 
\end{itemize} 
This $d$ dimensional Carrollian manifold can be described by a fibre bundle with the $(d-1)$ dimensional spatial directions forming the base space and and the temporal direction forming a $1d$ fibre is on top of this. Of this class of manifolds, we will be most interested in the flat Carrollian manifold: 
\begin{equation}
	\mathcal{C} = \underbrace{\mathcal{R}}_{fibre} \times \underbrace{\mathcal{R}^{d-1}}_{base} , \quad \tilde{h} = \delta_{ij}dx^i\otimes dx^j ,\quad \theta = \p_{t}
\end{equation} 
	Carroll Lie algebra is identified with those vector fields $\xi = \xi^a\p_{a}$ of $\mathcal{C}$, which satisfy the isometry conditions: 
	\begin{eqnarray}\label{CLa}
		\mathcal{L}_{\xi}\tilde{h}_{\mu\nu}=0,\quad \mathcal{L}_{\xi}\theta=0.
	\end{eqnarray}
	Solving these one obtains (for details the reader is pointed to Appendix \ref{CI})
	\begin{eqnarray}
		\xi^i = \omega^i_j x^j + b^i ,\qquad \xi^0 = a + f(x^k)
	\end{eqnarray}
	So the isometry group of Carroll structure is infinite-dimensional. Further requiring metric compatibility with the connection defined by $\nabla$, the function $f(x^k)$ is restricted to be linear and the Carroll algebra becomes finite dimensional and is given by \refb{carral}. To include Conformal Carroll isometries, one has to modify the isometry eq. $(\ref{CLa})$ to the condition for conformal isometries in the Carroll background: 
	\begin{eqnarray}
		\mathcal{L}_{\xi}\tilde{h}=\lambda_1 \tilde{h},\quad \mathcal{L}_{\xi}\theta= {\lambda_2}\theta.
	\end{eqnarray}
for some function $\lambda$ on $\mathcal{C}$. It can be shown that these conformal Carroll generators in $d$ dimensions close to form exactly the BMS$_{d+1}$ algebra \cite{Duval:2014uoa} when 
\be{}
N = - \frac{\lambda_1}{\lambda_2} = 2.
\ee
This specific value is the one for which space and time scale in the same way under the Dilatation generator. For other values of $N$, one gets more generic Carroll conformal algebras. 

\newpage	

\section{Carroll Clifford Algebra}\label{sec3}
We wish to build towards an understanding of fermions in Carrollian spacetimes. To this end, we first postulate a Carrollian version of the relativistic Clifford algebra and then we will discuss some aspects of its representation theory. Before we embark on this journey, let us quickly remind ourselves of relativistic fermions and the properties which we would need to construct Carroll analogues of. 

\subsection{Relativistic Fermions}
Fermions make up all the matter around us. The fundamental particles leptons, quarks are all fermions. Not only are fermions of central physical significance, their mathematical structure is also very rich. They live not on the group manifold that defines the spacetime but on the universal cover of it. One of the principal objects that define fermions is the Clifford algebra associated with it. 

\medskip

For relativistic fermions, the fundamental equation is of course the Dirac equation
\begin{equation}
(i\g^{\mu}\p_{\mu} - m)\Psi = 0,
\end{equation}
which can be derived from the corresponding Dirac action 
\begin{equation}\label{DA}
S = \int d^dx \sqrt{-g} \, \bar{\Psi}(i\g^{\mu}\p_{\mu} - m)\Psi.
\end{equation}
$\Psi$ is a four-component Dirac spinor and $\bar{\Psi}$ is defined as $\bar{\Psi} = \Psi^\dagger \g^0$. Here $\g$ matrices obey the Clifford algebra:
\begin{equation}\label{rca}
\{\g^{\mu},\g^{\nu}\}=-2g^{\mu\nu}. 
\end{equation}
where $g^{\mu\nu}$ is the spacetime metric whose determinant is given by $\sqrt{-g}$ and appears in the Dirac action \refb{DA}. Let us focus on $d=4$ and flat spacetimes $g^{\mu\nu}=\eta^{\mu\nu}= \text{diag}(-1,1,1,1)$. A particularly useful representation of the Clifford algebra here is given by: 
\begin{eqnarray}
\g^0 =  \begin{pmatrix} 0 & I \\ I & 0 \end{pmatrix}
\qquad	\g^i = \begin{pmatrix} \sg^i & 0\\ 0 & -\sg^i\end{pmatrix} ,\{i=1,2,3\}
\end{eqnarray}
with $\sg^i$ being the Pauli matrices. The gamma matrices obviously carry the information of the spacetime as is clear from the Clifford algebra. This becomes even more transparent when we link $\sg$ with the spacetime isometry algebra. 
Using the $\g$ matrices we can construct a set of matrices \begin{equation}
\Sigma^{\mu\nu}=\frac{1}{4}[\g^{\mu},\g^{\nu}].
\end{equation}
These matrices form a representation of Lorentz algebra
\begin{equation}\label{LA}
[\Sigma^{\mu\nu},\Sigma^{\rho\sigma}]=-\eta^{\mu\rho}\Sigma^{\nu\sigma}+\eta^{\nu\rho}\Sigma^{\mu\sigma}-\eta^{\nu\sigma}\Sigma^{\mu\rho}+\eta^{\mu\sigma}\Sigma^{\nu\rho}
\end{equation}
Here $\Sigma^{0i}$ is representation of Lorentz boost, whereas $\Sigma^{ij}$ is of $SO(d-1)$ rotation.	

\subsection{The Carrollian Algebra}
For Carrollian spacetimes, we have two metrics given by $\tilde{h}_{\mu\nu}$ and $\Theta^{\mu\nu}$. We need both to generate the isometry of Carroll spacetimes. In keeping with this we propose that the Carroll Clifford algebra is given by:  
\be{carcli}
{\boxed{\big\{\gc_{\mu},\gc_{\nu}\big\}=2\tilde{h}_{\mu\nu},\quad \big\{\gb^{\mu},\gb^{\nu}\big\}=2\Theta^{\mu\nu}}}
\ee
For flat Carroll spacetimes, which is the case we will be interested in throughout this work: 	
\be{}
\tilde{h}_{\mu\nu} = \text{diag}(0,1,1, \ldots) \quad \Theta^{\mu\nu} = \text{diag}(1,0,0,\ldots).
\ee
Contrary to the relativistic case, here we see that there are two sets of $\g$ matrices which obey different versions of the Carroll Clifford algebras. In principle, these could lead us to different types of fermions on a Carrollian manifold. In order to investigate if there is a preferred type of Clifford algebra or a preferred type of fermion, let us push on.  

\medskip

As a further test of our proposal, we will need to check that we can construct the set of matrices $\ts$ and $\hat{\Sigma}$ in analogy with \refb{LA}, which should now close to the equivalent of the Lorentz (and not Poincare) algebra. This is the sub-algebra of Carroll algebra which contains the Carroll boosts and rotations but excludes translation generators. For the lower gammas $\gc$, we construct 
\be{sc}
\ts_{ab} \equiv \frac{1}{4}[\gc_a, \gc_b]
\ee
and find 
\begin{subequations}\label{carr}
\begin{eqnarray}
&&[\ts_{0i},\ts_{0j}]=0, \quad [\ts_{0i},\ts_{jk}]=-\delta_{ij}\ts_{0k}+\delta_{ik}\ts_{0j}\\
&&[\ts_{ij},\ts_{kl}]=-\delta_{il}\ts_{jk}+\delta_{ik}\ts_{jl}-\delta_{jk}\ts_{il}+\delta_{jl}\ts_{ik}.
\end{eqnarray}
\end{subequations}
Here $\ts_{0i}$ are the Carroll boosts and $\ts_{ij}$ are spatial rotations. As seen from the above, the algebra does close to form the homogeneous part of the Carroll algebra. So, the lower gammas pass this test. 
\medskip

However, the same is not apparently true for the upper gammas. We again define 
\be{suc}
\hat{\Sigma}^{ab} \equiv \frac{1}{4}[\gb^a, \gb^b]. 
\ee
Although we get 
\be{}
[\hat{\Sigma}^{0i},\hat{\Sigma}^{0j}]=0
\ee
which means that the Carroll boosts are recovered, the rotation matrices are identically zero as the product of $\gb^i$ in Carroll case is zero:
\be{}
\hat{\Sigma}^{ij}=0.
\ee
The rotation-boost and the rotation-rotation commutators are hence identically zero and the $\hat{\Sigma}^{ab}$ do not form a faithful representation of the Carroll algebra in the background. We will present a way around this potential hurdle in Section \eqref{sec4}. Below we construct representations for both sets of gamma matrices, and later in the paper we construct actions for both sets of gamma matrices. 
	
\subsection{Representation of the Carroll Clifford algebra} 
In this subsection, we will construct an explicit representation of our Carroll Clifford algebra in $d=4$ and show how to get this from a lower dimensional representation. This will help us build representations for arbitrary dimensions. 
\subsubsection{Representation in $d=4$}
A representation of our Carroll Clifford algebra in $d=4$ flat Carroll spacetimes that would prove particularly important is the following: 
\begin{eqnarray}\label{lg4}
		\begin{split}
			\gc_0 = 
			\begin{pmatrix}
				0 & 0 & 0 \,\,\,& 0 \\
				0 & 0 & 0 & 0 \\
				i & 0 & 0 & 0 \\
				0 & \,\,i & 0 & 0
			\end{pmatrix},
 		\gc_1 =
			\begin{pmatrix}
				0 & 1 & 0 & 0 \\
				1 & 0 & 0 & 0 \\
				0 & 0 & 0 & -1 \\
				0 & 0 & -1 & 0
			\end{pmatrix},
		\gc_2 =
			\begin{pmatrix}
				0 & -i & 0 & 0 \\
				i & 0 & 0 & 0 \\
				0 & 0 & 0 & i \\
				0 & \,\,0 & -i & 0
			\end{pmatrix},
			\gc_3 =
			\begin{pmatrix}
				1 & 0 & 0 & 0 \\
				0 & -1 & 0 & 0 \\
				0 & 0 & -1 & 0 \\
				0 & 0 & 0 \,\,\,& 1
			\end{pmatrix}
		\end{split}
	\end{eqnarray}
	These matrices obey the following algebra
	\begin{equation}\label{lowcliff}
		\big\{\gc_{\mu},\gc_{\nu}\big\}=2\tilde{h}_{\mu\nu}.
	\end{equation}

Again following our arguments above, we can construct the $\ts$ matrices \refb{sc} and show that the Carrollian spacetime algebra \refb{carr} is satisfied by these $\ts$ matrices. This is thus a proper representation of the Carroll Clifford algebra.

\medskip

Now we construct a representation for the upper gammas. Here we will have another set of matrices given by
\begin{eqnarray}\label{ug4}
		\begin{split}
			\gb^0 = 
			\begin{pmatrix}
				1 & 0 & 0 & 0 \\
				0 & 1 & 0 & 0 \\
				0 & 0 & -1 & 0 \\
				0 & 0 & 0 & -1
			\end{pmatrix}
			\quad	\gb^1 =
			\begin{pmatrix}
				0 & 0 & 0 & 0 \\
				0 & 0 & 0 & 0 \\
				0 & i & 0 & 0 \\
				i & 0 & 0 & 0
			\end{pmatrix}
		\quad
			\gb^2 =
			\begin{pmatrix}
				0 & 0 & 0 & 0 \\
				0 & 0 & 0 & 0 \\
				0 & 1 & 0 & 0 \\
				-1 & 0 & 0 & 0
			\end{pmatrix}
			\quad	\gb^3 =
			\begin{pmatrix}
				0 & 0 & 0 & 0 \\
				0 & 0 & 0 & 0 \\
				i & 0 & 0 & 0 \\
				0 & -i & 0 & 0
			\end{pmatrix}
		\end{split}
	\end{eqnarray}
	It can be checked that these matrices obey the algebra 
	\begin{equation}\label{upcliff}
		\{\gb^{\mu},\gb^{\nu}\}=2 \Theta^{\mu\nu}.
	\end{equation}
As discussed before, the ${\hat{\Sigma}}^{ab}$ matrices constructed out of these $\gb^a$'s have the peculiar property that the rotation matrices ${\hat{\Sigma}}^{ij}$ vanish identically. The boosts ${\hat{\Sigma}}^{0i}$ are non-trivial and form an abelian algebra, as expected from Carrollian boosts. 
\medskip

{One must note again that the absence of rotation generators in the upper gamma representation can be traced back to their degenerate structure. However, we have not yet discussed the structure of discrete symmetries associated to these representations, and whether or not the above generators are compliant with those symmetries. In a later section we will ponder further over these discrete symmetries in $d=4$, and come back to a modified proper definition of Carroll rotation generators which will form a faithful representation of the algebra.  }
	
\subsubsection{Representation in general even dimensions}
Consider $d=2$. The degenerate Clifford algebra and its solutions were previously discussed in the context of tensionless superstrings \cite{Bagchi:2017cte}. We will borrow some of the nomenclature here.  Let us begin by considering the lower Carroll Clifford algebra in $d=2$ \refb{lowcliff}. This reduces to the matrix equations
	\begin{equation}
		(\gc_0)^2=0 ,\quad (\gc_1)^2=I ,\quad \gc_0\gc_1+\gc_1\gc_0=0.
	\end{equation}
A trivial solution is $\gc_0=0$ and $\gc_1=I$. This is what we will call the homogeneous case, keeping with the nomenclature borrowed from \cite{Bagchi:2017cte}. We will briefly comment on this later in the paper. The non-trivial solution has the form 
\begin{eqnarray}\label{inhom1}
\gc_0 = 
\begin{pmatrix}			
0 & 0 \\
a & 0
\end{pmatrix}
\qquad	\gc_1 =
\begin{pmatrix}
1 & 0 \\
0 & -1 
\end{pmatrix} 
\end{eqnarray}
with $a \in \mathcal{R}$. Now we will use these 2d matrices as seed matrices to generate generic higher dimensional matrices. The prescription we use to get the lower $\gc$ matrices for higher even dimensions is the following
\begin{eqnarray}\label{2tt4}
\hspace{-1cm} \gc_{'0}=\gc_{0} \otimes \begin{pmatrix} -1 & 0 \\0 & 1 \end{pmatrix}, \,
		\gc_{'i}=\gc_{i} \otimes \begin{pmatrix}
			-1 & 0 \\
			0 & 1
		\end{pmatrix}, \,
		\gc_{'d-2}=I \otimes \begin{pmatrix}
			0 & 1 \\
			1 & 0
		\end{pmatrix},\,
		\gc_{'d-1}=I \otimes \begin{pmatrix}
			0 & -i \\
			i & 0
		\end{pmatrix}.
	\end{eqnarray}
Here $\gc_{\mu}$ are the $2^k\times2^k$ matrices in $d-2$ dimensions and $I$ is the $2^k\times2^k$ identity matrices. 
It can be checked that with the construction outlined above, starting with the 2d gamma matrices \refb{inhom1} one reproduces the 4d gamma matrices \refb{lg4}.	

\medskip

Now, we come to the upper gammas. The upper Clifford algebra in $d=2$ \refb{upcliff} reduces to the matrix equations
\begin{equation}\label{ccu}
(\gb^0)^2=I ,\quad (\gb^1)^2=0 ,\quad \gb^0\gb^1+\gb^1\gb^0=0.
\end{equation}
Again we have a homogeneous (or trivial) solution is $\gb^0=I$ and $\gb^1=0$. The non-trivial or inhomogeneous solutions are of the form
\begin{eqnarray}\label{inhom2}
\gb^0 = \begin{pmatrix}
			1 & 0 \\
			0 & -1
		\end{pmatrix}
		\qquad	\gc^1 =
		\begin{pmatrix}
			0 & 0 \\
			a & 0 
		\end{pmatrix} 
\end{eqnarray}
with $a \in \mathbb{R}$. We can get higher dimensional upper $\gb$ matrices starting from $d=2$. In $d=2k+2$, $(k=1,2,...)$
\begin{eqnarray}\label{2t4}
\hspace{-1cm}  \gb^{'0}=\gb^{0} \otimes I, \, \gb^{'i}= \gb^{i} \otimes I, \, 
		\gb^{'d-2}=\begin{pmatrix}
			0 & 0 \\
			I & 0
		\end{pmatrix} \otimes \begin{pmatrix}
			0 & 1 \\
			1 & 0
		\end{pmatrix}, \, 
		\gb^{'d-1}=\begin{pmatrix}
			0 & 0 \\
			I & 0
		\end{pmatrix} \otimes \begin{pmatrix}
			0 & -i \\
			i & 0
		\end{pmatrix}.
	\end{eqnarray}
Note that the construction for moving upwards in dimension for the lower and the upper gamma matrices are not identical. It can again be checked that with the prescription outlined the upper $2d$ gammas \refb{inhom2} become the upper $4d$ gammas we presented earlier in \refb{ug4}.

\subsubsection{Representation in general odd dimensions}
Having understood how to construct higher even dimensional representations from the 2d representation, we now address how to construct general odd dimensional representations for our deformed Clifford algebra \refb{carcli}. For this, we will explicitly construct a three dimensional representation and use our previous method of generating higher dimensional representations \refb{2tt4}-\refb{2t4}.  	
In $d=3$, one of the representations which obey lower Carroll Clifford algebra \refb{lowcliff} is given by
\begin{eqnarray}\label{3dg}
		\begin{split}
			\gc_0 = 
			\begin{pmatrix}
				0 & 0 & 0 \,\,\,& 0 \\
				0 & 0 & 0 & 0 \\
				i & 0 & 0 & 0 \\
				0 & \,\,i & 0 & 0
			\end{pmatrix}
			\qquad	\gc_1 =
			\begin{pmatrix}
				0 & 1 & 0 & 0 \\
				1 & 0 & 0 & 0 \\
				0 & 0 & 0 & -1 \\
				0 & 0 & -1 & 0
			\end{pmatrix}
	\qquad
			\gc_2 =
			\begin{pmatrix}
				0 & -i & 0 & 0 \\
				i & 0 & 0 & 0 \\
				0 & 0 & 0 & i \\
				0 & \,\,0 & -i & 0
			\end{pmatrix}.
			\end{split}
\end{eqnarray}
Notice that unlike the relativistic case, we use a higher $4\times4$ dimensional representation for these matrices. 
In order to generate higher odd dimensional matrices, we use these 3d matrices as seed matrices and then follow the method outlined earlier in \refb{2tt4}. For example, going from 3d to 5d, the representation matrices will be constructed such as 
	\begin{eqnarray}\label{2ttt4}
		\gc^{'}_{0}=\gc_{0} \otimes \begin{pmatrix}
			-1 & 0 \\
			0 & 1
		\end{pmatrix} \quad 
		\gc^{'}_{i}=\gc_{i} \otimes \begin{pmatrix}
			-1 & 0 \\
			0 & 1
		\end{pmatrix} \quad 
		\gc^{'}_{3}=I \otimes \begin{pmatrix}
			0 & 1 \\
			1 & 0
		\end{pmatrix}\quad 
		\gc^{'}_{4}=I \otimes \begin{pmatrix}
			0 & -i \\
			i & 0
		\end{pmatrix}.
	\end{eqnarray}
Here $\gc_0,\gc_{i}$ are the matrices in 3d representation $(\ref{3dg})$, and $I$ is $4\times 4$ matrix. 

\medskip

For the upper Clifford algebra \refb{upcliff}, in $d=3$, we get the following matrices:
\begin{eqnarray}
\gb^0 = 
		\begin{pmatrix}
			1 & 0 & 0 & 0 \\
			0 & 1 & 0 & 0 \\
			0 & 0 & -1 & 0 \\
			0 & 0 & 0 & -1
		\end{pmatrix}
		\qquad	\gc^1 =
		\begin{pmatrix}
			0 & 0 & 0 & 0 \\
			0 & 0 & 0 & 0 \\
			0 & i & 0 & 0 \\
			i & 0 & 0 & 0
		\end{pmatrix}
	\qquad
		\gb^2 =
		\begin{pmatrix}
			0 & 0 & 0 & 0 \\
			0 & 0 & 0 & 0 \\
			0 & 1 & 0 & 0 \\
			-1 & 0 & 0 & 0
		\end{pmatrix}.
	\end{eqnarray}
These can be used along with eq $(\ref{2t4})$ to construct representations for higher odd dimensional upper Carroll gamma matrices. 

\subsection{Structure of Carroll spinors in $d=4$}
There are further representation dependent structures in a theory of spinors, which form the basis of discrete symmetries. Since we will be moving on to discuss the structure of Carroll invariant fermion actions, it will be instructive to clear up the nature of some of these symmetries for the Carroll case. It would turn out that the structures discussed here provide a solution to the apparent problem with the upper gammas not providing a faithful representation of the background Carroll algebra. 

\medskip

We first start with spinor $\Psi$ and wish to define its adjoint, like we do for the Lorentz invariant case. The reader is reminded that relativistic Dirac adjoints are usually defined as $\overline\Psi = \Psi^\dagger \gamma^0$, and these lead to an array of bilinears that change covariantly under Lorentz transformations that map spinors according to $\Psi\rightarrow S(\hs)\Psi$.  Now, since Carroll Clifford algebra is manifestly degenerate, defining an adjoint with $\g^0$ would be problematic. We thus introduce an analogue of the Dirac adjoint:  
	\begin{equation}\label{adj1}
		\bar{\Psi} = \Psi^\dagger \Lambda, \quad \text{where} \quad \Lambda = 
		\begin{pmatrix}
			0 & iI \\
			-iI & 0
		\end{pmatrix}
	\end{equation} 
	such that $\Lambda^2=I$. The matrix $\Lambda$ is defined so that the term $\bar{\Psi}\Psi$ behaves like a scalar under Carroll transformations.  To justify this choice, note that hermitian conjugate of the upper $\g$  matrices now satisfy
\be{}
\gb^{\mu^\dagger}=-\Lambda \gb^{\mu} \Lambda^{-1}
\ee
which implies the Carroll invariant condition for the generators,
\be{}
	\hs^{\mu\nu\dagger}=-\Lambda \hs^{\mu\nu} \Lambda^{-1} \implies S(\hs)^{\dagger}
= \Lambda S(\hs)^{-1}  \Lambda, 
\ee
which in turn keeps the mass terms invariant under Carroll transformations. Note that, for representation with lower $\g$ matrices one can argue the existence of a similar adjoint structure in the same vein. In fact, a short calculation leads us to the adjoint for lower gamma matrices as:
\be{adj2}
\Lambda= \begin{pmatrix}
			0 & I \\
			I & 0
		\end{pmatrix},
\ee

which is still idempotent. We will be using these definitions to write down our spinor actions throughout the paper. 
\medskip

We now focus on the discussion about the structure of Charge Conjugation symmetries in the Carroll case. In general for the Lorentz case, one would like to work with real fermions that stay real under Lorentz boosts, and hence a Majorana condition $\Psi^* = \Psi$ is imposed on the spinors at every spacetime point. In a generic Clifford basis, one would define the charge conjugate as: 
\be{}
\Psi \mapsto \Psi ^{c}=\,\mathcal{C}{\bar {\Psi }}^{{T}},
\ee
where $\mathcal{C}$ is a $4\times 4$ matrix satisfying 
\begin{eqnarray}\label{}
		\quad\,\gb^{\mu^T} = -\mathcal{C}\gb^{\mu}\mathcal{C}^{-1},~~\mathcal{C}^\dagger \mathcal{C} = \mathbf{I}.
\end{eqnarray}	
One can easily check this is a good definition, i.e. the conjugate spinors transforms accordingly under Carroll transformations. A viable solution to the above conditions can be written here as:
\begin{eqnarray}\label{chrgc}
	\mathcal{C} = 
	\begin{pmatrix}
		0 & i\sigma^2 \\
		i\sigma^2 & 0
 	\end{pmatrix}
\end{eqnarray}	
where it can be further checked that
	\begin{eqnarray}
		\hs^{\mu\nu^T} = -\mathcal{C}\hs^{\mu\nu}\mathcal{C}^{-1},\quad 
		\gb^{\mu^*} = {(\Lambda\mathcal{C})}\gb^{\mu}(\Lambda\mathcal{C})^{-1},\quad
		\hs^{\mu\nu^*} = {(\Lambda\mathcal{C})}\hs^{\mu\nu}(\Lambda\mathcal{C})^{-1}.
	\end{eqnarray}
Once we put this structure in place, we can go ahead and define a Majorana-like condition for Carroll spinors which simply reads $\Psi^c = \Psi$. One can similarly find an equivalent representation for charge conjugation corresponding to the lower gamma matrices, however they would not be very important for our purpose here. 

\newpage

\section{Action for Carroll fermions and its symmetries}\label{sec4}
Equipped with the basic structures and required tools, we now want to construct Carroll invariant actions for fermions specialising to the case for $d=4$. In this section, we will start with a two component Carroll bispinor, and write down actions in the Dirac form involving both lower and upper gamma matrices. 

\medskip

As we have seen before, although lower gamma matrices can be used to construct perfectly faithful representations of the Carroll algebra, the upper gammas fail in this regard due to vanishing rotation generators. We will remedy this via demanding the rotation generators to be compliant with charge conjugation symmetries, and this leads to a general faithful representation even for the upper gamma theory. We will discuss the explicit invariance of these actions under Carroll symmetries.  

\medskip

Once that is done, we would present a general Carroll invariant action in 4d by just demanding all individual terms to be Carroll boost and rotation invariant, and show these invariant terms encompass both lower gamma action and the modified upper gamma action.

\medskip

Finally, we will give a physical meaning to the two different varieties of Carroll fermions arising out of the lower and upper gammas. 

\subsection{Action with lower gammas}
We will begin by looking at an action constructed out of the lower gamma matrices in \eqref{lg4}. Since the $\ts$ matrices constructed out of these gammas give a proper representation of the spacetime Carroll algebra (minus translations), we expect that the action constructed out of these matrices would have invariance under the full Carroll group and when the mass terms are turned off, we expect to see conformal Carroll invariance appearing. 
\medskip
%Let us take the hermitian conjugate of the $\gc_\mu$  matrices for this representation: 
%\be{}
%\gc_{\mu}^\dagger=-\Lambda \gc_{\mu} \Lambda \quad \text{which implies} \quad \ts_{\mu\nu}^\dagger=-\Lambda \ts_{\mu\nu} \Lambda
%\ee	

Evidently, due the faithful representation furnished by the the $\gamma_{\mu}$ matrices, we have $\bar{\Psi}\gc_{0}\p_{t}\Psi $ and $\bar{\Psi} \Psi$ as two Carroll invariant terms, corresponding to the kinetic and the mass term for a theory of a 4-component fermion $\Psi$. Here adjoint spinors are defined as \eqref{adj2}. Hence the following action is a good point to start from:
	\begin{equation}\label{CL}
		S_{\text{lower}} = \int dt d^3x \big(\bar{\Psi}\gc_{0}\p_{t}\Psi -m\bar{\Psi}\Psi\big). 
	\end{equation}
One can also write a second type of mass term here given by 
\be{Wmass}
\mathcal{L}_{\text{mass}} = m\bar{\Psi}\gc_0\Psi.
\ee
It is easily checked that this mass term is indeed Carroll invariant. We will use this second mass term in the discussion in terms of the two-component spinors below, although our main interest lies in the massless or conformal theory. 

\medskip

Due to the degeneracy in the $\gamma$ matrices displayed above, only two of the four components of the Carroll-Dirac fermion $(\Psi)$: 
\begin{eqnarray}
	\Psi = \begin{pmatrix}
		\phi\\
		\chi
	\end{pmatrix}
\end{eqnarray}
take part in the dynamics and the action, now with second type of mass term, boils down to:
\begin{equation} \label{2complag}
S_{\text{lower}} = \int dt d^3x \left(i \phi^\dagger \p_{t}\phi-m  \phi^\dagger \phi \right),
\end{equation}
where $\phi$ is a 2-component spinor.
% One may wish to further write $\phi$ in terms of component Grassmann fields: $$\begin{pmatrix}
%\chi_1 \\
%\chi_2 
%\end{pmatrix},$$ so that the action \eqref{2complag} simplifies to:
%\begin{eqnarray}
%S_{\text{lower}} = \int dt d^3x   \sum_{l=1}^{2}\left(i \chi^{\star}_i \p_t \chi_i -  m  \chi^{\star}_i \chi_i  \right)
%\end{eqnarray}
For the massive case, this enjoys a $SU(2)$ symmetry at each space point, which enlarges to $SO(4)$ for the massless case.  Even for the global parts of these transformations, these are richer than the relativistic case, where one encounters $U(1) \times U(1)$ symmetry only for Weyl (massless fermions) and just a $U(1)$ for the massive one. 
\medskip

We recall that due to the degeneracy in the representation of the Carrollian Clifford algebra, we have a theory of two component spinors, resulting into the dynamics of a single degree of freedom. Moreover the components of the spinor $\phi$ are decoupled in \eqref{2complag}, indicating the said representation to be reducible. Hence the minimalistic Carrollian fermion theory should have half degree of freedom \footnote{See, eg. chapter seven of \cite{Henneaux:1992ig}.}. 

\subsection*{Continuous symmetries}	
For $m \neq 0$, the free theory \eqref{2complag} enjoys space-time translation, spatial rotation and the infinite number of supertranslation symmetries. Rotation acting on the two component spinor $\phi$ can be easily read off-from the 4 dimensional representation of the $\Sigma_{ij}$ matrices as:
	\begin{eqnarray} \label{rot_spin_half}
		&&\delta_{n} \phi = \epsilon^{ijk} n_i x_j \partial_k \phi - \frac{i}{2} n^l \sigma_l \phi.%+i\epsilon_{ijk}\sigma^k
%		\nonumber \\
%		&& \delta_b \phi = b_i x^i\p_{t}\phi \nonumber \\
%		&& \delta_{f} \phi = f(x) \p_{t}\phi.
	\end{eqnarray}
for a constant $3$ vector $n^i$.	
On the other hand the supertranslations act on the 2-component spinor $\phi$ respectively as:
\begin{eqnarray}
	&& \delta_{f} \phi = f(x) \p_{t}\phi.
	\end{eqnarray}
%\textcolor{red}{Question: Here we are proposing $\phi$ as spin 1/2 as per $\gamma$ matrix commutators. However the action is invariant even if we treat $\phi$ as spin 0. It now points out that two component version of $\phi$ is redundant. Qn. What is the fundamental building block of Carroll fermions?}	
	
Here by $f$ we denote any tensor field on $\mathbb{R}^3$. The special cases $f =1, f = x^i $ and $x^i x_i$	respectively correspond to time translation, Carrollian boost and the temporal part of the special conformal transformation. It is to be noted that although the mass term breaks special conformal symmetry, the temporal part of it still holds to be a symmetry. Also note that, since the components of $\phi$ are decoupled, the spin matrix doesn't appear in the spatial rotation generator.
\medskip

For the massless case, as expected, the dilatation and the spatial part of the special conformal transformation are symmetries as well:
\begin{eqnarray}
		&&\delta_{\Delta}\phi=(t\p_{t}+x_{k}\p_{k}+\Delta)\phi,  \nonumber \\
		&& \delta_{k}\phi=k^j \left( 2\Delta x_{j}+2x_{j}t\p_{t}+2x_{i}x_{j}\p_{i}\phi-x_{i}x_{i}\p_{j} -i \epsilon_{ijk} x^i \sigma_k \right)\phi.
	\end{eqnarray} 

For this case invariance under dilatation imposes that $\Delta=\frac{d-1}{2}$. For $d=4$, $\Delta=\frac{3}{2}$, as in the relativistic case. This matches with the previous exposition with Carrollian fermions \cite{Bagchi:2019xfx}. It is interesting to note here that even though the whole finite CCA does not turn out to be symmetries of the massive system, due to the obvious breaking of scaling invariance by the mass term, the infinite dimensional supertranslations remain a symmetry of \refb{CL}, even with this mass term. 
\medskip

It's time we discuss the internal symmetries of the 2-component theory of Carrollian fermions \eqref{2complag}.
%\subsection{Local $U(1)$ symmetries}
Rewriting the 2-component Grassman field, stripped to the real parts as:
$$ \phi = \begin{pmatrix}
\phi_1 + i \phi_2 \\
\phi_3 + i \phi_4
\end{pmatrix}, $$
we see the that the action \eqref{2complag} becomes:
\begin{eqnarray}
S_{\text{lower}} = i \int dt d^3x\,\left( \sum_{l=1}^4 \phi_l \p_t \phi_l - 2 m \left(\phi_1 \phi_2 + \phi_3 \phi_4 \right) \right) .
\end{eqnarray}
The kinetic part is manifestly invariant under $SO(4)$ symmetry transformations, with arbitrary spatial dependence, which breaks to $SU(2)$ for the mass term. The corresponding Noether charges respectively for the massless and the massive case are:
\begin{eqnarray} \label{charges}
Q_{(m=0)} = \int d^3 x \, \lambda^{ij} (x) \phi_i \phi_j, ~~~Q_{(m\neq 0)} = \int d^3 x \, \phi^{\dagger} \mathbf{n}(x) \cdot \pmb{\sigma}  \phi .
\end{eqnarray}
Here $\lambda$ are $\mathbf{R}^3$ valued lie algebra elements of $\mathfrak{so}(4)$ and $\mathbf{n}(x)$ are vector fields on $\mathbf{R}^3$.
\medskip

We discuss a rather intriguing feature of these lower Carroll fermion before ending this subsection regarding the space-time rotation symmetry of the action \eqref{2complag}. It is evident that the spin-matrix part appearing in the rotation transformation rule \eqref{rot_spin_half} is derived from the Clifford algebra representation ($\Sigma_{ij}$), substantiating $\phi$ as a spin-1/2 field. However, the action \eqref{2complag}, having now spin mixing involved is invariant, even if we treat $\phi$ as a scalar (spin-0) under space-time rotation. This indicates that Carroll symmetry doesn't necessitate fermions (anti-commuting fields) to have half-integral spin representations. We will give a more intuitive picture of why this peculiar feature arises here at the end of this section. 

%As the free theory \eqref{2complag} is ultra-local in space, it's evident that the global $U(1)$ symmetry can be promoted to local ones, even without being necessitated by gauge fields:
%\begin{eqnarray}
%\Psi\rightarrow e^{i\lambda(x)}\phi%,\quad \bar{\Psi}\rightarrow e^{-i\lambda(x)}\bar{\Psi}
%\end{eqnarray}
%The conserved charges are
%\begin{eqnarray}
%Q_{\lambda}=\int d^3 x \, \phi^{\dagger}\phi \lambda
%\end{eqnarray}
%and they form a center-less abelian affine algebra:
%\begin{eqnarray}
%\{ Q_{\lambda_1}, Q_{\lambda_2} \} = \delta_{\lambda_2} Q_{\lambda_1} = 0
%\end{eqnarray}

\subsection*{The Hamiltonian}
Before we go on to actions for upper gammas, let us point out another interesting feature of this action that will be of interest later in the paper. The momentum conjugate to the spinor is given by
\begin{eqnarray}
	\Pi = \p \mathcal{L}/\p \dot{\Psi} = \bar{\Psi}\gc_{0}.
\end{eqnarray}
The Hamiltonian for the Carroll fermions is given by 
\be{lowerH}
\mathcal{H}_{\text{lower}}=\Pi\dot{\Psi} - \mathcal{L} = m  \phi^\dagger \phi,
\ee
where we have used the Weyl mass term \refb{Wmass}. We note that the Hamiltonian vanishes for the massless case. 

\medskip

The vanishing of the Hamiltonian in the massless case is also borne out by the computation of the energy-momentum tensor for the action. Starting from the Lagrangian \refb{CL}, one can use Noether's procedure to arrive at the energy-momentum tensor for these lower Carroll fermions: 
\begin{equation}
	T^{a}_{\,\,\,\ b} = \frac{\p L}{\p(\p_{a}\Psi)}\p_{b}\Psi  + \frac{\p L}{\p(\p_{a}\bar{\Psi})}\p_{b}\bar{\Psi} - \delta^{a}_{\,\, b}\mathcal{L}
\end{equation}
Component-wise the above turns out to be: 
\begin{eqnarray}
T^t_{\,\,\,t} =  m  \phi^\dagger \phi, \quad T^i_{\,\,\,t} = 0, \quad T^t_{\,\,\,i} = i \phi^\dagger \p_{i}\phi, \quad T^i_{\,\,\,j}= -\delta^i_j \mathcal{L}.
\end{eqnarray}
The vanishing of $T^i_{\,\,\,t}$ is a tell-tale sign of a Carroll invariant system, arising from Carroll boost invariance. The further vanishing of $T^t_{\,\,\,t}$, in the massless case, indicates the consequent vanishing of the Hamiltonian of the system. Using equations of motion, it is also obvious that the stress tensor defined is traceless for the massless case indicating the emergence of conformal Carroll symmetry in the system.

\subsection{Action with upper gammas}
We now consider upper gamma matrices to construct our action. Since the sigma matrices arising out of these have the peculiar feature that the ones corresponding to rotations identically vanish, we can see actions written from here seemingly will not be invariant under the whole Carroll group. 
\medskip

We start with an action in the Dirac form as in the following, again concentrating on the $4d$ case: 
	\begin{equation}\label{D1}
S_{\text{upper}}= \int dt d^{3}x \, \bar{\Psi}(\gb^{0}\p_{t}+\gb^{i}\p_{i})\Psi.
	\end{equation}
	Note that there is no $i$ in this Lagrangian to make it Hermitian with the choice of $\Lambda$ in \eqref{adj1}.
	Here as we have described before, $(\gb^{0})^2=1$ and $(\gb^{i})^2=0$. For the moment, we will only consider the massless case. Under boost transformation the above action is invariant but  we can clearly see that the vanishing of the $\hs^{ij}$ in this representation of upper gammas leads to a variation of the action which is not seemingly rotation invariant, as all brackets containing $\hs^{ij}$ simply drop out. For the special case of $d=2$, however, since there are no rotation generators, this action becomes invariant under the whole Conformal Carroll group. Interestingly, the symmetries get infinitely enhanced to the full BMS$_3$ algebra \refb{}. The details of this and other important implication of Carroll fermions in $d=2$ are addressed in a companion paper \cite{2dcarroll}. 
	\medskip

However, all is not lost for the upper gammas. As we commented earlier, this apparent non-invariance stems from the peculiar degenerate structure of the spatial $\gb^i$ matrices. Remember, those do not have a clear hermitian structure. Hence, we need to find a generator for the $SO(3)$ subalgebra which is manifestly compliant with the charge conjugation symmetries.  In this regard, let us define a modified rotation generator that makes the spin current symmetric under charge conjugation:
\begin{eqnarray}
	%&&\ts^{0i} = \frac{1}{4}[\gc^0,\gc^i]\\
	&&\hs^{ij}_c = \frac{1}{4}[\gb^i,\gb^{j}_c] + \frac{1}{4}[\gb^{i}_c,\gb^j]
\end{eqnarray}
where we have put in a subscript $c$ to remind the reader of the fact that the charge conjugation plays a crucial role in making up these modified rotation generators. Here 
\be{}
\gb^i_c = -\mathcal{C}\gb^i \mathcal{C}^{-1},
\ee 
and $\mathcal{C}$ is as given in \eqref{chrgc}. What this effectively does is to impose proper Hermiticity onto the rotation subalgebra, which was erstwhile missing. Note here that spatial lower gammas $\gc_i$ are not degenerate and hence do not need this modified definition. As a matter of fact, in any representation where the Hermitian conjugate of spatial gamma matrices are related to itself upto a sign (relativistic Dirac matrices for example), the above definition boils down to the usual definition of rotation generators. Using these above representation the generators of Carroll algebra become matrices of the form:
\begin{eqnarray}
		\hs^{0i}_c = -\frac{1}{2}
		\begin{pmatrix}
			0 & 0 \\
			\sigma^i & 0
		\end{pmatrix}
		 ,\quad
		\hs^{ij}_c = \frac{i}{2}\epsilon^{ijk}
		\begin{pmatrix}
			\sigma^k & 0 \\
			0 & \sigma^k
		\end{pmatrix}.
	\end{eqnarray}
They can be checked to explicitly close the Carroll boost-rotation algebra,$\quad$
\bes
\begin{eqnarray}
		&&[\hs^{0i}_c,\hs^{0j}_c]=0	\\
		&&[\hs^{0i}_c,\hs^{jk}_c]=-\delta^{ik}\hs^{0j}_c+\delta^{ij}\hs^{0k}_c\\
		&&[\hs^{ij}_c,\hs^{kl}_c]=\delta^{il}\hs^{jk}_c-\delta^{ik}\hs^{jl}_c+\delta^{jk}\hs^{il}_c-\delta^{jl}\hs^{ik}_c.
	\end{eqnarray}
	\ees
Also this furthermore implies
\begin{eqnarray}
	&&(\hs^{0i}_c)^\dagger = -\Lambda\hs^{0i}_c\Lambda, \quad (\hs^{ij}_c)^\dagger = -\Lambda\hs^{ij}_c\Lambda.
\end{eqnarray}
Equipped with a now proper faithful representation of the Carroll generators, we can now go ahead to investigate the symmetries of our massless Dirac upper gamma action in \eqref{D1}.

\subsection*{Continuous symmetries}
Of course the action is trivially invariant under translations. In what follows let us mention the non-trivial symmetries of the total bispinor $\Psi$. Under the action of Carroll boosts, the spinors and their adjoints change as:
\begin{eqnarray}
	\delta_C \Psi = (x_i\p_{t}-\hs^{0i}_c)\Psi, \quad \delta_C \bar{\Psi} = (x_i\p_{t}\bar{\Psi}+\bar{\Psi}\hs^{0i}_c)
	\end{eqnarray} 
	then the action is invariant under these changes as the Lagrangian variation $\delta_C \mathcal{L} = \p_t(x^i\mathcal{L})$ is a total derivative{\footnote{The reader may be a bit perturbed by the structure of the indices in the above equations, but we reminder her that we are not thinking of a Carroll covariant formulation in this work. The $\hs_c$ are just the rotation matrices constructed out of the upper gammas, taking into account the charge conjugation.}}. Under rotations, the fields change as
	\begin{eqnarray}
	\delta_{J}\Psi=(x_{i}\p_{j}-x_{j}\p_{i}-\hs^{ij}_c)\Psi, \quad 
	\delta_{J}\bar{\Psi}=(x_{i}\p_{j}-x_{j}\p_{i})\bar{\Psi}+\bar{\Psi}\hs^{ij}_c.
	\end{eqnarray}
	Remember this was the non trivial one in the case of rotation generators beforehand. Since the modified generator and commutators involving it do not vanish, one can see that the change in the Lagrangian would be 
	\begin{eqnarray}
		\delta_J \mathcal{L} = \p_{j}\big(x_{i}\mathcal{L}\big)-\p_{i}\big(x_{j}\mathcal{L}\big),
	\end{eqnarray}
	and hence, invariance will be restored. One can go further ahead and check for all global generators of the Carroll Conformal Algebra in $4d$, including:\\
	
	\textit{Dilatation}: 
	\be{}
		\delta_{D}\Psi=(t\p_{t}+x_{k}\p_{k}+\Delta)\Psi, ~~~
		\delta_{D}\bar{\Psi}=(t\p_{t}+x_{k}\p_{k}+\Delta)\bar{\Psi}
	\ee
	\textit{Temporal Special Conformal Transformation}:
	\be{}
	\delta_{K}\Psi=-x_{i}x_{i}\p_{t}\Psi  + 2x_i\hs^{0i}_c\Psi~~~
		\delta_{K}\bar{\Psi}=-x_{i}x_{i}\p_{t}\bar{\Psi} - 2x_i\bar{\Psi}\hs^{0i}_c.
		\ee
	\textit{Spatial Special Conformal Transformation}:
	\bes \begin{eqnarray}&&
		\delta_{K_{j}}\Psi=2\Delta x_{j}\Psi+2x_{j}t\p_{t}\Psi+2x_{i}x_{j}\p_{i}\Psi-x_{i}x_{i}\p_{j}\Psi - 2x_i\hs^{ij}_c\Psi - 2t\hs^{0j}_c\Psi\\&&
		\delta_{K_{j}}\bar{\Psi}=2\Delta x_{j}\Psi+2x_{j}t\p_{t}\bar{\Psi}+2x_{i}x_{j}\p_{i}\bar{\Psi}-x_{i}x_{i}\p_{j}\bar{\Psi} + 2x_i\bar{\Psi}\hs^{ij}_c + 2t\bar{\Psi}\hs^{0j}_c .
	\end{eqnarray} \ees
	One can show clear invariance of the Lagrangian under all these finite CCA transformations. Specifically, under Dilatations, the spinors change again with $\Delta = \frac{3}{2}$ as in the case before.
	
	\subsection*{Hamiltonian and Stress Tensor}
	
	In the case of Lower gamma representations, we were concerned with only one two component block of the bispinor. Since in the current case all four components are involved in the dynamics, let us clearly define our spinor structure:
	\begin{eqnarray}
	\Psi = \begin{pmatrix}
		\phi\\
		\chi
	\end{pmatrix}
\end{eqnarray}
where both $\phi,\chi$ are two component spinors. With this structure, the lagrangian \eqref{D1} can be written in terms of components:
 \begin{equation}\label{upga}
 	\mathcal{L}_{\text{upper}} = -i\big(\chi^\dagger\dot{\phi} + \phi^\dagger\dot{\chi} - \phi^\dagger\sigma^i\p_i\phi  \big),
 \end{equation}
with a clear presence of space derivatives unlike the lower gamma case. One can compare this action with the $2d$ worldsheet actions found in relation to Inhomogeneous super-BMS symmetries for tensionless strings in \cite{Bagchi:2017cte, Bagchi:2018wsn} and see this is a natural higher dimensional analogue of the one written there. Now we can easily find out the equations of motion: 
\begin{eqnarray}
	\dot{\phi}=0,\quad \dot{\chi} = \sigma^i\p_i\phi.
\end{eqnarray}
Noting the conjugate momenta $\Pi_{\Psi} = \bar{\Psi}\gc^0$, it is then also easy to construct the Hamiltonian density:
\begin{eqnarray}
	\mathcal{H}_{\text{upper}} = \Pi\dot{\Psi} - \mathcal{L}_{\text{upper}} = -\bar{\Psi}\gb^i\p_i\Psi = i\phi^\dagger \s^i\p_i\phi.
\end{eqnarray}
Again, note that contrary to the lower gamma case, the massless Hamiltonian does not vanish. 

Starting from the Lagrangian using the Noether's procedure we can calculate the energy-momentum tensor for upper Carroll fermions:
\begin{equation}
	T^{\mu}_{\,\,\,\nu} = \frac{\p L}{\p(\p_{\mu}\Psi)}\p_{\nu}\Psi  + \frac{\p L}{\p(\p_{\mu}\bar{\Psi})}\p_{\nu}\bar{\Psi} - \delta^{\mu}_{\,\,\,\nu}\mathcal{L}.
\end{equation}
Here we can write component wise,
\begin{eqnarray}
	T^t_{\,\,\,t} = -\bar{\Psi}\gb^i\p_{i}\Psi, \quad T^i_{\,\,\,t} = \bar{\Psi}\gb^i\p_{t}\Psi, \quad T^t_{\,\,\,i} = \bar{\Psi}\gb^{0}\p_{i}\Psi, \quad T^i_{\,\,\,j}= \bar{\Psi}\gb^i\p_{j}\Psi-\delta^i_j \mathcal{L}.
\end{eqnarray}

Trace of the EM tensor $T^{\mu}_{\,\,\,\mu} = -3\mathcal{L}$ which vanishes using equation of motion. Futhermore, we can see $T^i_{\,\,\,t}$ vanishes on-shell for this case, signifying the onset of Carroll physics. 

\medskip

An indication of emergent Carrollian symmetry in a system is the vanishing of the commutator of Hamiltonians at different spacetime points \cite{Henneaux:2021yzg}. Although in the case of the lower fermions, this was trivial, due to the presence of spatial derivatives, this is no longer obvious for the upper fermions. We check this explicitly below. One can impose the canonical commutation relations, which read
\begin{eqnarray}
	&&\{\Pi(x),\Psi(x')\} = \delta(x-x') \rightarrow \{\bar{\Psi}(x),\Psi(x')\} = \gb^0\delta(x-x') \nonumber\\
	&&\{\Psi(x),\Psi(x')\} = \{\bar{\Psi}(x),\bar{\Psi}(x')\} = 0
\end{eqnarray}
As a further consistency check of Carroll invariance, one can try to evaluate the commutator between Hamiltonian densities,
\begin{eqnarray}
	[\mathcal{H}(x),\mathcal{H}(x')] &&= [\bar{\Psi}(x)\gb^i\p_i\Psi(x),\bar{\Psi}(x')\gb^j\p_j\Psi(x')]\nonumber\\
	&&=\bar{\Psi}(x)\gb^i\gb^0\gb^j\p_i\delta(x-x')\p_j\Psi(x')-\bar{\Psi}(x')\gb^j\gb^0\gb^i\p_j\delta(x-x')\p_i\Psi(x)\nonumber\\
	&&=-\bar{\Psi}(x)\gb^0\gb^i\gb^j\p_i\delta(x-x')\p_j\Psi(x')+\bar{\Psi}(x')\gb^0\gb^j\gb^i\p_j\delta(x-x')\p_i\Psi(x) \nonumber
\end{eqnarray}
Now using Clifford algebra , the product of two $\gb^i$ matrices is identically zero. So the above commutator vanishes:
\begin{equation}
[\mathcal{H}(x),\mathcal{H}(x')] = 0.
\end{equation}

\subsection{Generic Carroll boost invariant action}
As the reader noticed, in the last section we modified the definition of rotation generators to retrace the invariance of our upper gamma action under $SO(3)$ subalgebra. Since this may have come in as a novelty, and the seeming non-invariance of the case earlier was puzzling enough, let us give some more arguments in favour of our upper gamma action. 
\medskip

We set out to find the most general $4d$ spinor lagrangian invariant under Carroll boosts are rotations. 
In terms of two 2-component spinors $\phi$ and $\chi$, one can write the most generic Lagrangian involving all possible temporal and spatial derivative terms as
\begin{equation}
	\mathcal{L} = a\phi^\dagger\dot{\phi} + b\chi^\dagger\dot{\chi} + c\phi^\dagger\dot{\chi} + d\chi^\dagger\dot{\phi} + e\phi^\dagger\sigma\cdot\p \phi + f\chi^\dagger\sigma\cdot\p \chi + g\phi^\dagger\sigma\cdot\p \chi + h\chi^\dagger\sigma\cdot\p \phi
\end{equation}
Here all the coefficients are in general complex.
The two component spinors transform under Carroll boost and rotation as
\begin{eqnarray}
&&\delta_b \phi = b^ix_i\p_t\phi, ~~~~
\delta_b \chi = b^i(x_i\p_t\chi + \alpha \sigma_i\phi) \\
&&\delta_n \phi = \epsilon^{ijk}n_kx_i\p_j\phi + \frac{i}{2}n^i\sigma_i\phi, ~~~~
\delta_n \chi = \epsilon^{ijk}n_kx_i\p_j\chi + \frac{i}{2}n^i\sigma_i\chi .
\end{eqnarray}
Note that $\phi$ and $\chi$ mix under boosts for the transformation of $\chi$. The parameter $\a$ comes from a ambiguity in the Clifford Algebra, since a rescaling of $\gb^i$ matrices do not change the algebra. This constant rescaling parameter appears in the spin-matrix part of the boost transformation rule. The effect of this (the `$a$' in \eqref{inhom2}) is very interesting in the $2d$ Conformal Carroll case, where this scaling gives rise to an automorphism of the superrotation algebra \cite{2dcarroll}.
Invariance of the test lagrangian under these symmetries imply certain constraints on the constant coefficients:  
\begin{eqnarray}
 c\alpha + d\alpha^* + e =0, \quad b=f=g=h=0.
\end{eqnarray}
Hence the form of boost-rotation invariant Lagrangian gets modified as
\begin{equation}\label{genboost}
	\mathcal{L} = a\phi^\dagger\dot{\phi} + c\phi^\dagger\dot{\chi} + d\chi^\dagger\dot{\phi} + e\phi^\dagger\sigma\cdot\p \phi .
\end{equation}
We have to also ensure that the Lagrangian is Hermitian. This requires further constraints on the parameters:
\begin{eqnarray}
	a^* = -a,\, c = -d^*,\, e^* = -e.
\end{eqnarray}
Then without any loss of generality, we could choose $c= d =-i, ~a=e=i$, and the boost ambiguity can be chosen as $\a = \frac{1}{2}$ consistently. Now with this choice, we can clearly see our most general invariant Lagrangian \eqref{genboost} splits into a chiral part depending on $\phi$, which turns out to be the massless lower gamma action \eqref{2complag}, and the rest is simply identified with the upper gamma cousin of it \eqref{upga}. So, in general, any Carroll invariant fermionic Lagrangian can be written as
\begin{equation}
	\mathcal{L} = a_L \mathcal{L}_{\text{lower}} + a_U \mathcal{L}_{\text{upper}},
\end{equation}
where we have introduced two constants $a_L$ and $a_U$. We have not written any term that introduces an interaction between these two types of fermions. It would be interesting to see if such terms are allowed from geometric considerations of generic Carroll manifolds. 

\subsection{Physics of two types of Carroll fermions}
Let us pause to summarise our findings so far. The geometry of Carrollian manifolds suggested that there may be different types of fermions residing here. This was due to the two different Clifford algebras one could write down \refb{}. Following an analysis of the representation theory, in this section, we wrote down two distinct actions for the two different types of fermions \refb{} and \refb{}, which we will keep calling lower and upper. These led to theories, both Carroll invariant, but with different distinguishing features, e.g. a zero and a non-zero Hamiltonian. So what are these two different types of fermions, and can we say something more fundamental about them? Below we try to paint a more physical picture of what has so far been a mostly mathematical exercise.

\begin{figure}[!h]
\centering
\includegraphics[width=.7\linewidth]{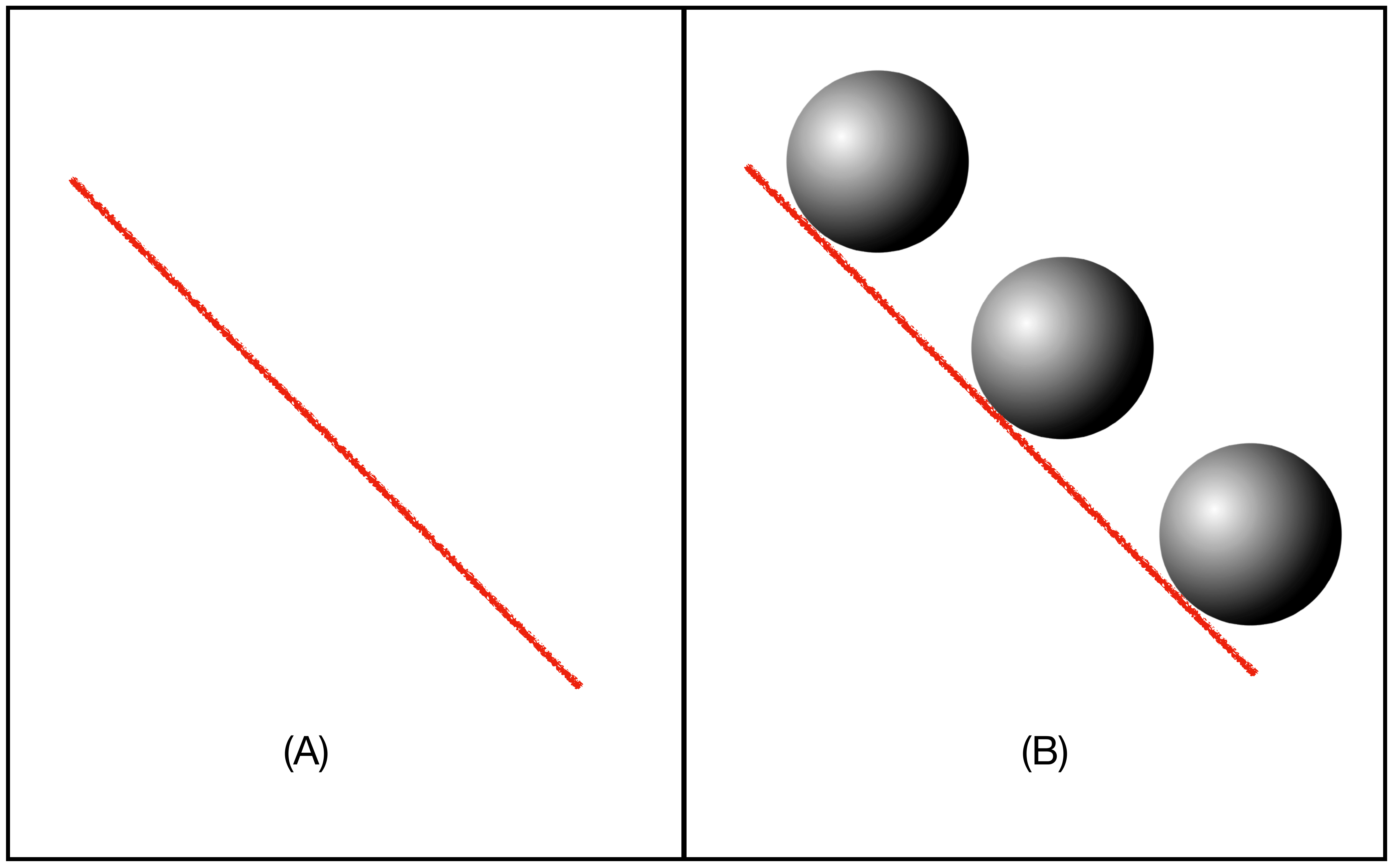}
\caption{Carroll fermions viewing $\mathscr{I}^\pm$: (A) Lower fermions, (B) Upper fermions}
\label{st-ferm}
\end{figure}

\medskip

\paragraph{Lower fermions:} The lower fermions come from the ``time-like" part of the Clifford algebra. This Clifford algebra does not know about what is happening in the spatial part of the Carroll spacetime. The action \refb{} also suggests that the spatial parts just go for a ride here. This is most apparent in the feature discussed below \refb{charges}. These lower fermions, although anti-commuting, need not have half-integral spins. This just means that the spin part, which resides in the $(d-1)$ dimensional spatial fibre of the $d$ dimensional Carrollian spacetime, does not have a defining role for the lower fermions. In the closed up lightcone, they reside purely along the null-timelike direction. In a sense, they behave like a one-dimensional theory living only on the null line, oblivious to the spatial fibre. 
%In two dimension, we can show that this leads to a chiral truncated fermion \cite{}. 
E.g. when we consider lower fermions on the null boundary of 4d Minkowski spacetimes $\mathscr{I}^\pm$, which is given by R$_u \times$S$^2$, these fermions would only see the null line parametrized by the retarded time $u$ and fail to see the celestial spheres attached to each point on the null line. This is depicted by part (A) in Figure \refb{st-ferm}.

\paragraph{Upper fermions:} The upper fermions, on the other hand, come from the ``spatial" part of the Carroll Clifford algebra. In the action \refb{} there are now spatial derivatives along with the temporal ones. The upper fermions clearly see the base of the fibre bundle, i.e. the null time direction, as well as the spatial fibres. These fermions are more usual, more like their relativistic counterparts. Their anticommuting nature comes hand in hand with their half-integral spins. Going back to the example of the null boundary of 4d flat spacetimes, the upper fermions see the whole R$_u \times$S$^2$ structure, as shown in part (B) in Figure \refb{st-ferm}.

\bigskip

The fact that the massless version of lower fermions have a vanishing Hamiltonian and the upper massless fermions have a non-vanishing one and they behave in very different ways can also be traced back to the representation theory of the Carroll algebra, where the Hamiltonian is a central element and the two cases $H=0$ and $H\neq0$ leads to different classes of representations. See \cite{deBoer:2021jej} for more details on these classes of representations. 

\bigskip

Carroll theories, generically, are of two types, Electric and Magnetic. These names have been derived from the Galilean counterparts and specifically the theory of electromagnetism in the non-relativistic world where one can have two distinct limits called the electric and magnetic limits where the electric effects dominate over magnetic ones and vice versa. Similar phenomena have been seen in generic field theories in the Carrollian regime, and the electric theories are defined as ones where temporal features dominate over spatial ones and similarly for magnetic theories. Even Carroll scalars come in these forms {\footnote{There is even a third type of Carroll scalar in two spacetime dimensions \cite{Bagchi:2022eav}.}}. From our above discussion, it seems that lower fermions are Electric and the upper ones are Magnetic. 

\medskip

In generic field theories, the electric theory is obtained as the leading term in a $c\to0$ limit of the corresponding relativistic theory, while the magnetic theory comes from the sub-leading term. A naive analysis of the $c\to0$ limit of a relativistic fermionic field theory runs into some hurdles which we have not immediately been able to sort out. We hope to return to this in the near future. 

\newpage

\section{Flat Bands and the Emergence of Carroll}\label{sec5}

The relation between energy and momentum, the so-called dispersion relation, is at the heart of many complicated phenomena in real life condensed matter systems. For non-relativistic free particles, energy depends on the square of the spatial momentum and this simple fact coupled with the requirement of a minimum energy state leads to bosons bunching together and forming a Bose-Einstein condensate at zero momentum. Fermions on the other hand avoid each other and fill up the dispersion curve up to the Fermi energy. This simple quadratic dispersion explains a lot of the physics of metals, semi-conductors and simple insulators. Low temperature properties of magnetic materials are understood in terms of linear or quadratic dispersion relations. 

\medskip

Of late, there has been a surge of interest in understanding flat band physics, where the energy don't depend on momentum at all. Usual intuitions don't hold in this regime, --- bosons don't condensate, fermions don't form usual Fermi surfaces. A lot of the known physics breaks down. Flat band physics seems appear in some very interesting systems including spin liquids,  fractional quantum hall systems, and twisted bi-layer graphene, where superconductivity appears at unexpected temperatures. 

\subsection{A quick argument}

Below we will go on to show that Carroll invariance is a generic feature of such flat band physics and then give a flavour of this in the twisted bi-layer graphene system. But before we proceed with the calculations and explicit demonstrations, here is a quick physical argument of why one should expect Carroll invariance to appear in these contexts. 

\medskip

We are attempting to understand flat bands. This is a limit where the cone in position space with the energy and momentum collapses onto the momentum plane. A priori this looks like a non-relativistic limit where the speed of light goes to infinity and the lightcone opens up completely. But we must remember that we are in momentum space. Opening up of the cone in momentum space is equivalent to the actual position space lightcone closing down. This is obviously the Carrollian limit and hence whenever one encounters flat bands in systems, one will generically end up with Carrollian structures {\footnote{This understanding came about in a quick discussion with Blagoje Oblak. AB thanks him for asking the question which led to this picture.}}. 
\medskip

\begin{figure}[!h]
\centering
\includegraphics[width=.9\linewidth]{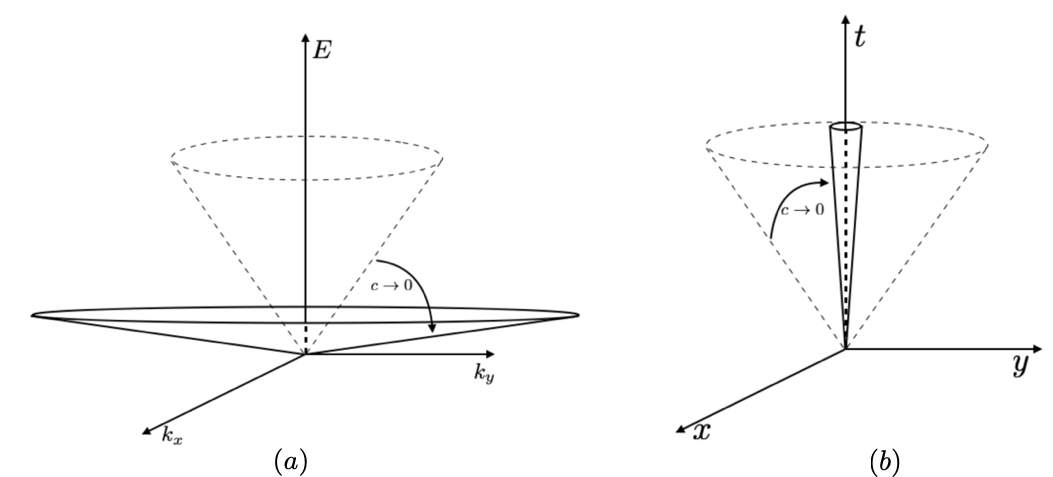}
\caption{(a) Flattening out of the momentum space dispersion relation and (b) closing down of space-time light cone $c \rightarrow 0$.}
\label{collapse}
\end{figure}
Our discussion above and explicit calculations below also tie up very nicely with the claims in the condensed matter literature that flat bands generically lead to localisation in position space. From a Carroll perspective, this localisation in space is a manifest consequence of the closing up of the lightcone as the speed of light goes to zero. Having understood this very quick intuitive and physical picture, we make the emergence of Carroll symmetry very explicit below. 

\subsection{Emergence of Carroll} \label{connections}
Let us consider the following free theory Hamiltonian, expressed in terms of infinite number of Heisenberg picture oscillators (assume the system to be autonomous):
\begin{eqnarray}\label{first_master}
H = \sum_n E_n a^{\dagger}_n a_n ,
\end{eqnarray}
where we will take the liberty of  considering the oscillators to be uncountably infinite as well, thus being able to convert the sum to integral when focusing on specific models. Also the index $n$ should be taken as a mnemonic for a number of indices, denoting other charges present in the system. 
\medskip

Let us start with the case of fermions, $i.e.$ equal-time anti-commuting oscillators 
\be{}
\{ a^{\dagger}_n, a_m\} = \delta_{mn}
\ee 
 where it is understood that the Kronecker delta is to be replaced by Dirac delta as and when needed. Now we make an assumption that there exists an operator $\mathfrak{h}$, which is self adjoint on the Hilbert space of square integrable (generalizable to multi-component/ vector-values, as in the case of Weyl or Dirac spinors) functions on $\mathbb{R}^d$, with suitable boundary conditions. Also we demand that the eigenvalues of $\mathfrak{h}$ are $E_n$ as appearing in \eqref{first_master}. $u_n(\mathbf{r})$ are orthonormal eigenfunctions: 
\be{}
\mathfrak{h} u_n(\mathbf{r}) = E_n u_n(\mathbf{r}).
\ee 
Hence the Hamiltonian can be repackaged as:
\begin{eqnarray} \label{repack}
&& H = \sum_{m,n} \int d^d\mathbf{r}\, a^{\dagger}_m(t) u^{\star}_m (\mathbf{r}) \mathfrak{h} \left(a_n(t) u_n (\mathbf{r}) \right) = \int d^d\mathbf{r} \, \Psi^{\dagger} (\mathbf{r},t )\mathfrak{h} \Psi (\mathbf{r},t ) \\
&& \mbox{where } \Psi (\mathbf{r},t ) := \sum_n a_n(t) u_n(\mathbf{r}). \nonumber
\end{eqnarray}
The definition of the field operator $\Psi$ above, in terms of the oscillators directly imply the following non-zero anti-commutator:
\begin{eqnarray}
\{ \Psi (\mathbf{r},t) , \Psi^{\dagger}(\mathbf{r}',t) \} = \delta^d (\mathbf{r} - \mathbf{r}').
\end{eqnarray}
At the classical level, identifying $\Psi^{\star}$ as the canonically conjugate momentum of $\Psi$, the action can be written via Legendre transformation \footnote{This involves a second class constraint, which is easily solved. Also, for $\mathfrak{h}$ being a matrix-valued differential operator, ie, $u$ being multi-component functions, the inner product needs to be adjusted with insertion of another matrix for the sake of self-adjointness.}:
\begin{eqnarray}
S = \int dt\, d^d \mathbf{r} \Psi^{\star} \left( i\partial_t - \mathfrak{h} \right) \Psi .
\end{eqnarray}
As is well known, for $d=3$ Weyl fermions, the dispersion in \eqref{first_master} is $E_{\mathbf{k}} = c |\mathbf{k}|$, while for massive Dirac fermions, $E_{\mathbf{k}} = c \left( \mathbf{k}^2+ m^2 c^2 \right)^{1/2}$ and $\mathfrak{h}$ is 2 and 4 dimensional matrix valued differential operator respectively for these cases. The large $c$ limit of the later gives us: $E_{\mathbf{k}} \approx \frac{\mathbf{k}^2}{2m}$, which describes the Schr{\"o}dinger field theory and is described by $\mathfrak{h} = -\frac{1}{2m} \nabla^2$, on single component fermions.
%Finally consider the dispersionless free theory of fermions, described by the Hamiltonian
%\begin{eqnarray}
%H= E  \sum_n  a^{\dagger}_n a_n .
%\end{eqnarray}

\medskip

In line of the above discussion, let's for simplicity take a single component continuum field $\Psi$ to describe the real space degrees of freedom. We now impose the supertranslation transformations on it as:
\begin{eqnarray}
\delta_{\mathbf{f}} \Psi(\mathbf{r},t) = f(\mathbf{r}) \partial_t \Psi,
\end{eqnarray}
for any real function $f$ on $\mathbb{R}^d$. As described earlier in this paper, $f=1$ is the time translation and $f= \mathbf{b} \cdot \mathbf{r}$, for any $d$-vector $\mathbf{b}$ is the Carrollian boost. We don't have an inhomogeneous term because this is a single component case. On the space of Heisenberg picture oscillators 
\be{}
a_m(t) = e^{-iE_m \,t} a_m (0)
\ee 
this transformation acts as:
\begin{eqnarray} \label{a_tran}
\delta_f a_m (t) = -i  \sum_{n} E_n \int d^d \mathbf{r}\, f(\mathbf{r}) \, u_n (\mathbf{r})  u^{\star}_m (\mathbf{r}) a_n(t) 
\end{eqnarray}
and similarly on $a^{\dagger}_m$. The transformation of the fundamental anti-commutators is
\begin{eqnarray}
\delta_f \{a_m, a^{\dagger}_n \} = - i \int d^d \mathbf{r} \,f(\mathbf{r}) u_n (\mathbf{r})  u^{\star}_m (\mathbf{r}) (E_n -E_m)
\end{eqnarray}
Note that the above expression vanishes identically for $f =1$ due to orthonormality of the eigenfunctions $u_n$. This is expected, since by definition the system is time translation invariant.  A sufficient condition for the above to vanish for an arbitrary $f$ is  $$E_n = E_m \quad \forall \, m,n,$$ 
 ie, the dispersion is trivial. Or in other words, the Hamiltonian has to be a flat-band one. The invariance of the brackets $\{a_m, a_n\}$ and $\{a^{\dagger}_m, a^{\dagger}_n\}$ does not put any more conditions on the dispersion. Furthermore, with the flat-dispersion Hamiltonian $H = E \sum_{n} a^{\dagger}_n a_n$ also is invariant under the super-translation transformations:
\begin{eqnarray}
\delta_f H  = 0.%&=& i \sum_{n,m} \int d^d \mathbf{r} \, f(\mathbf{r})\,E_n \left(  u^{\star}_n (\mathbf{r})  u_m (\mathbf{r}) a^{\dagger}_n a_m - u_n (\mathbf{r})  u^{\star}_m (\mathbf{r})a^{\dagger}_m a_n \right) \nonumber\\
%&=& 
\end{eqnarray}
The canonical structure and the Hamiltonian for the flat-band system remaining invariant implies that the system enjoys infinite number of Carrollian supertranslation symmetries.

Since $E_n$ are the eigenvalues of the differential operator $\mathfrak{h}$, one comes across a flat-dispersion situation whenever  $\mathfrak{h}$ doesn't contain any derivative at all and only is a constant. This is certainly true if the spatial manifold is $\mathbb{R}^d$ and one has open boundary conditions. However there exists non-trivial $\mathfrak{h}$ containing derivatives operating on Hilbert space of functions, whose domains are compact submanifolds of $\mathbb{R}^d$ with periodic boundary conditions. We will illustrate a couple of such cases in what follows. 

\medskip

A further caveat is that, for a multi-component $\Psi$, in general the $\mathfrak{h}$ dispersion set comes in the form of a number of disconnected branches/ bands, some of which may cross at a finite number of points. Since the distinct bands describe dynamics of decoupled degrees of freedom, for any band that is pristine and flat (i.e. not crossing any other bands), the corresponding degrees of freedom will have Carrollian supertranslations as symmetry. Moreover, such a pristine flat band, if it is the lowest energy one, would capture effective quantum dynamics from the Wilsonian perspective. 

\medskip

Before delving into examples emerging from low energy physics, let's try to understand connection between flat-bands and Carroll symmetry starting from Lorentz covariant systems having trivial open boundary conditions. The Hamiltonian of a free massive Lorentz covariant theory has the structure, in terms of fourier basis oscillator modes:
\begin{eqnarray} \label{master}
H = \int d^d \mathbf{k} \sqrt{c^2 \mathbf{k}^2 + m^2 c^4} \, a^{\dagger}_{\mathbf{k}} a_{\mathbf{k}} .
\end{eqnarray}
We have just one component and suppressed the spin and polarization information for convenience. The dispersion relation $E_{k} = \sqrt{c^2 \mathbf{k}^2 + m^2 c^4}$ bears the fact that the theory is Lorentz covariant. Written in real space, the integrand in the Hamiltonian will involve spatial derivatives in fields, resulting in to non-commuting Hamiltonian densities. But, a direct consequence of Carroll invariance is commuting Hamiltonian densities , as per \cite{Henneaux:2021yzg}. Therefore it follows that theories with no spatial derivative in real space Hamiltonian and hence having a non-dispersive Hamiltonian in terms of Fourier modes should have Carrollian symmetry. We have also seen that there are Carrollian theories having spatial derivatives, like in the case of our upper fermions. But still the Hamiltonian densities commute and there is emergent Carrollian symmetry. 

\medskip

One straightforward way to construct such a theory is starting with relativistic dispersion $E_\mathbf{k}= \sqrt{c^2 \mathbf{k}^2 + m^2 c^4}$ and take $c \rightarrow 0$ with $m c^2 = E_0$ remaining finite. $E_0$ defines the UV cut-off at which the theory is defined. Hence $E_k = E_0$ becomes dispersion-less (or flat-band, as commonly referred to in condensed matter physics literature). For the special case of $m=0$, the dispersionless Hamiltonian is identically zero.

\newpage

\section{Carroll Fermions in Condensed Matter Systems}\label{sec6}
In this section, we look at two examples of Carrollian fermions in real life condensed matter systems. Our first example will be that of a $1d$ chain, while the second example be in $2+1$ dimensions and will involve twisted bi-layered graphene, which shows emergent superconductivity. We will give some details of usual graphene and its bi-layer cousin for the benefit of people like ourselves who may be unfamiliar with these systems.

\subsection{Carroll fermions in one dimensional chain}
We begin in $1+1$ dimensions, with a 1 dimensional lattice model of fermions known as the diamond network ladder \cite{leykam2018artificial} or the Creutz ladder \cite{creutz2001aspects}. Consider, as per the figure below, a ladder consisting of a couple of parallel 1 dimensional crystals (arbitrarily long), each having same translational lattice symmetry with lattice parameter $A$. In the end of the analysis, we will take $A\rightarrow 0$.  As per our model, for each chain, tightly bound electrons can experience nearest neighbour hopping with energy $t$. Also there are inter-chain hoppings, which are both nearest ($\gamma$) as well as next-to-nearest neighbour ($\gamma'$).
\begin{figure}[!h]
\centering
\includegraphics[width=.7\linewidth]{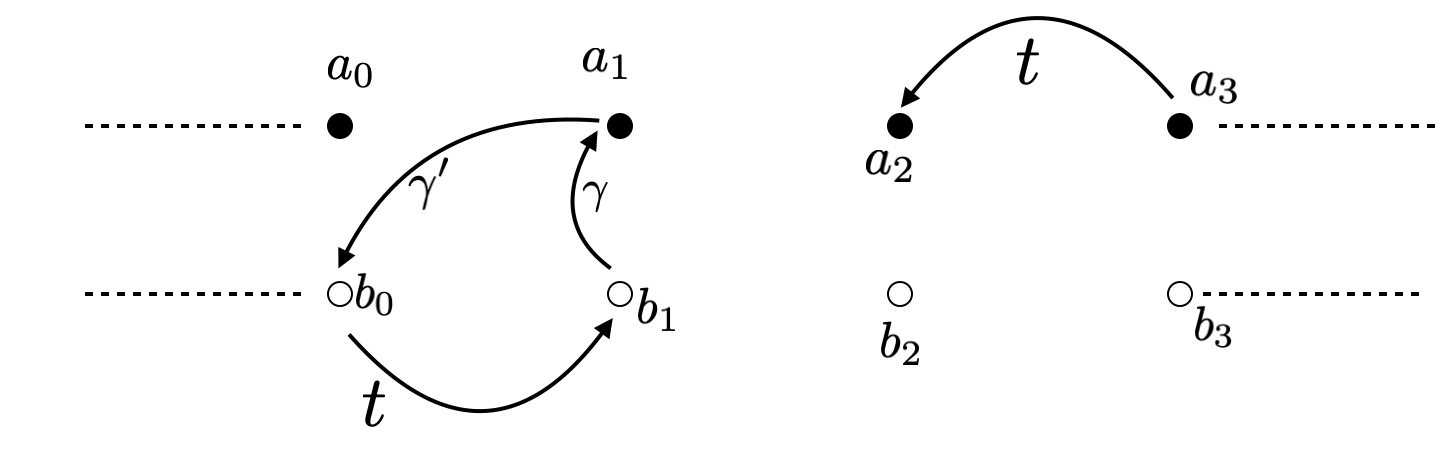}
\caption{The coupled chain system with all the intra-chain and inter-chain hopping strengths shown.}
\label{chain}
\end{figure}
The Hamiltonian in real space oscillator basis is:
\begin{eqnarray} \label{1Dham}
H =\sum_{p} t (a^{\dagger}_{p+1} a_p + b^{\dagger}_{p+1} b_p  ) + \gamma a^{\dagger}_{p} b_p  + \gamma' ( a^{\dagger}_{p+1} b_p + a^{\dagger}_{p} b_{p+1} ) + \mathrm{h.c.}
\end{eqnarray}
For periodic boundary conditions, one may go to the Fourier basis defined by 
\be{}
c_k = \sum_{p} a_p e^{ikAp}, d_k = \sum_{p} b_p e^{ikAp} \quad \text{for} \quad k\in (-\pi/A , \pi/A).
\ee 
As a consequence, the Hamiltonian becomes:
\begin{eqnarray} \label{kernel_creutz}
H =\sum_k  \begin{pmatrix}
c_k^{\dagger} \\
d_k^{\dagger}
\end{pmatrix} \begin{pmatrix}
2t\, \cos(kA) & \gamma + 2 \gamma' \cos(kA) \\
\gamma + 2 \gamma' \cos(kA) & 2t\, \cos(kA) 
\end{pmatrix} \begin{pmatrix}
c_k & d_k
\end{pmatrix}.
\end{eqnarray}
Diagonalizing, we get two bands:
\be{}
E^{\pm} (k) = 2 (t \pm \gamma') \cos(k) \pm \gamma.
\ee 
We readily observe that for $t= \gamma'$, $E^-$ becomes independent of $k$ and hence a flat-band appears. Hence for $t= \gamma'$, in another basis of modes, in which the matrix in \eqref{kernel_creutz} is diagonal, we have the Hamiltonian \eqref{1Dham} $H = H_{\mathrm{flat}} + H'$, with
\be{}
H_{\mathrm{flat}} = - \gamma \sum_k \tilde{c}^{\dagger}_k \tilde{c}_k ~~~\mbox{ and } ~~~~ H' = \sum_k  \left(4 t \cos(k) + \gamma \right)\tilde{d}^{\dagger}_k \tilde{d}_k
\ee
Now with the choice of $\gamma > 2t$, the bands don't cross and the flat-band Hamiltonian contains the lowest energy modes.
\begin{figure}[!h]
\centering
\includegraphics[width=.7\linewidth]{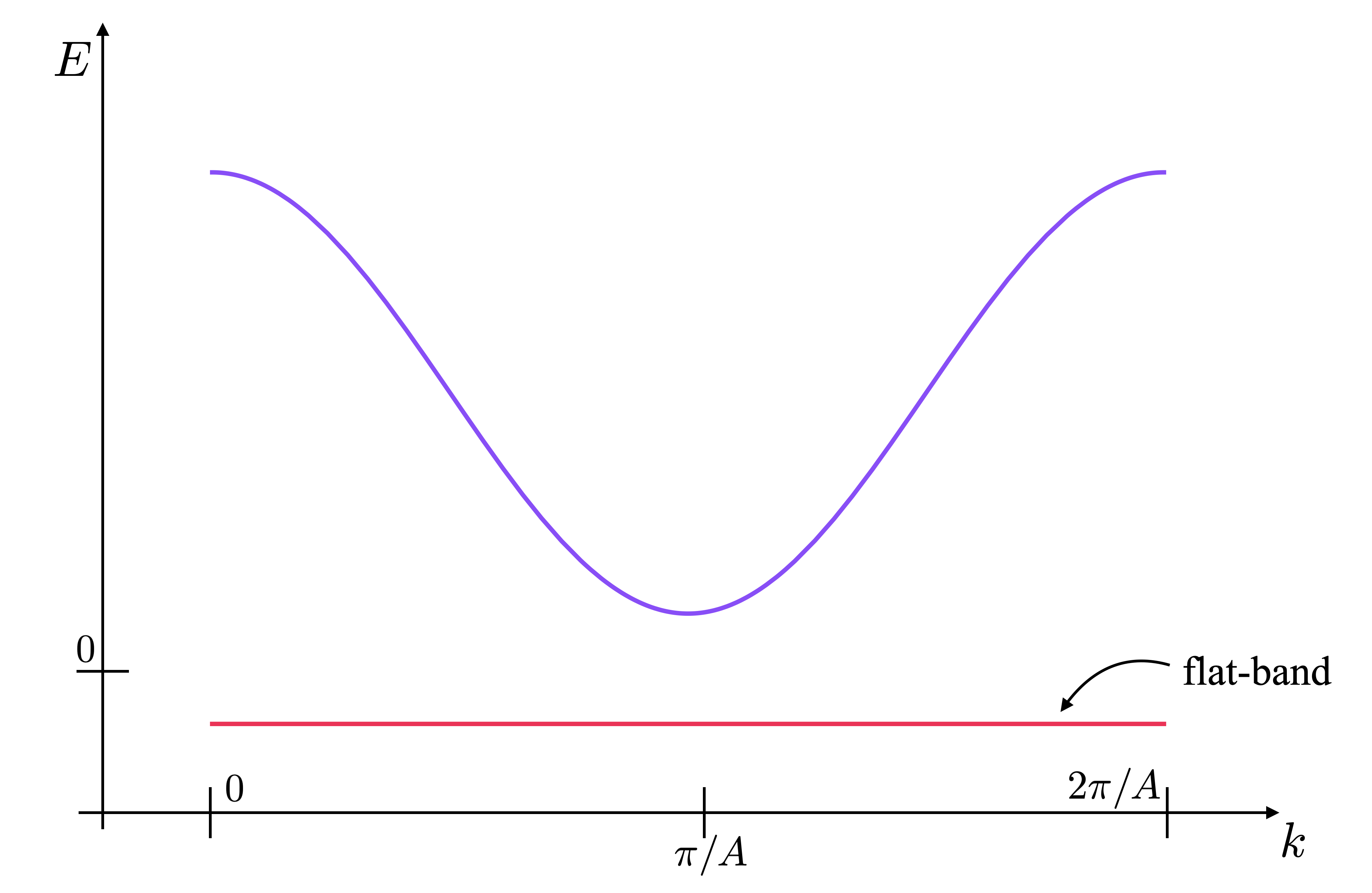}
\caption{The two dispersion bands, with the lower one being perfectly flat, for $\gamma' = \gamma > 2t$}.
\label{chain}
\end{figure}
Since the modes of $H_{\mathrm{flat}}$ contains the lowest lying modes, the effective dynamics is driven by it. In the continuum limit, as $A \rightarrow 0$, the Brillouin zone ($k \in [-\pi/A,  \pi/A]$) covers the whole of real line and $\tilde{c}_k \rightarrow \tilde{c}(k)$. Defining the fourier transform of it as the single component fermionic field $\psi(x)$, we have the Hamiltonian describing the low energy dynamics as:
\begin{eqnarray}
H_{\mathrm{flat}} = - \gamma \int dx\, \psi^{\dagger} \psi,
\end{eqnarray}
$\gamma$ playing the role of the mass term.

\medskip

This above Hamiltonian is very clearly the Hamiltonian for our lower fermions \refb{lowerH}. We have thus identified the physics of this condensed matter system to the formal Carrollian structures we had previously discussed. 

\subsection{Magic angle bilayer graphene}
Our next example would be a more involved one, that of bi-layer graphene twisted at particular angles called magic angles, where superconductivity appears. 

\medskip

Before delving in to the topic of bilayer graphene, let us remind ourselves that the effective low energy description of electrons in monolayer graphene is both scale (massless) and Lorentz invariant, with the Hamiltonian describing the free dynamics of two component spinor $\psi$:
\begin{eqnarray} \label{mlham}
H= i v_F \int d^2 \mathbf{r} \, \Psi^{\dagger} (\mathbf{r},t) \,   \pmb{\sigma}\cdot  \pmb{\nabla}\, \Psi(\mathbf{r},t),
\end{eqnarray}
giving rise to linear dispersion $E( \mathbf{k}) = v_F |\mathbf{k}|$. Here we keep in mind that the Fermi velocity $v_F \approx 3 \times 10^6 m/s$ plays the role of speed of light, but is 3 order in magnitude smaller than $c$ and is determined by the product of the hopping energy $\tau$ and the lattice parameter $a$. Also, the effective description \eqref{mlham}  and the space-time symmetries enjoyed by it are valid for energy scales $\ll 1/a$.

\medskip

When one stacks a couple of graphene layers on top of each other with a twist by an angle $\theta$, one needs to take into account interlayer couplings, which extends modifies the above monoloayer Hamiltonian \eqref{mlham}. The low energy continuum model capturing these features and robustly accounting for dispersionless states is the so-called `chiral' one \cite{tarnopolsky2019origin}:
\begin{eqnarray}
\label{blg_ham}
\tilde{H} = v_F \int d^2 \mathbf{r} \, \Phi^{\dagger} (\mathbf{r},t) D \Phi  (\mathbf{r},t).
\end{eqnarray}
Here, $\Phi^{\dagger} = \begin{pmatrix} \psi_1^{\dagger} & \psi_2^{\dagger} & \chi_1^{\dagger} &  \chi_2^{\dagger} \end{pmatrix} $ is a 4-component spinor, containing 2 for each $(\psi, \chi)$ graphene layer and $D$ is the matrix differential operator: 
\begin{eqnarray}
\label{Dop}
&& D =\begin{pmatrix} 0  &  \mathcal{D}^{\ast}( -\mathbf{r}) \\
  \mathcal{D}( \mathbf{r}) & 0 \end{pmatrix}, ~~  \mbox{ where } ~~ \mathcal{D}( \mathbf{r})  = \begin{pmatrix}
 -2i \bar{\partial} & \alpha U(\mathbf{r}) \\
 \alpha U(-\mathbf{r}) & -2i \bar{\partial}
 \end{pmatrix} \mbox{ and }\\
&& U(\mathbf{r}) = \sum_{j=0}^{2} e^{i\left( \frac{2\pi j}{3} - \mathbf{q}_j \cdot \mathbf{r}\right)} \mbox{ for }
  \mathbf{q}_0 = \frac{8\pi}{3a} \sin(\theta/2) (0,-1), \mathbf{q}_{1,2} =  \frac{8\pi}{3a} \sin(\theta/2) (\pm \sqrt{3}/2, 1/2) \non
\end{eqnarray}
%\begin{eqnarray}
%\label{Dop}
%&& D = \begin{pmatrix} -i\, v_0 \, \pmb{\sigma}_{\theta/2}\cdot \pmb{\nabla} & T(\mathbf{r}) \\
%T^{\dagger}(\mathbf{r}) & -i\, v_0 \, \pmb{\sigma}_{-\theta/2}\cdot \pmb{\nabla} \end{pmatrix}, \quad \mbox{where }\,  \pmb{\sigma}_{\theta/2} = e^{- i \sigma_z \theta/4} (\sigma_x, \sigma_y)e^{  i \sigma_z \theta/4}, \\
%&& T(\mathbf{r}) = w_1 \begin{pmatrix} 0 & e^{-i\mathbf{p}_1 \cdot \mathbf{r}}+e^{-i\left(\frac{2\pi}{3}+\mathbf{p}_2 \cdot \mathbf{r}\right)} +e^{-i\left(\frac{4\pi}{3}+\mathbf{p}_3 \cdot \mathbf{r}\right)} \\
%e^{-i\mathbf{p}_1 \cdot \mathbf{r}}+e^{i\left(\frac{2\pi}{3}-\mathbf{p}_2 \cdot \mathbf{r}\right)} +e^{i\left(\frac{4\pi}{3}-\mathbf{p}_3 \cdot \mathbf{r}\right)} &0
%\end{pmatrix} \non \\
 %\mbox{where } && \mathbf{p}_1 = \frac{8\pi}{3a} \sin(\theta/2) (0,-1), \mathbf{p}_{2,3} =  \frac{8\pi}{3a} \sin(\theta/2) (\pm \sqrt{3}/2, 1/2) \non
%\end{eqnarray}
%and $a$ is the lattice constant.
The dimensionless parameter $2 \pi \alpha = \frac{w_1}{\tau} \csc(\theta/2) $ controls the coupling between the layers through the interlayer hopping strength $w_1$. Any two of the $\mathbf{q}$ appearing in the equation above serve as the basis vectors in the `mini'-Brillouin zone. Note that, in the commensurate Moire pattern formed by the two mono-layer graphene sheets, the real space periodicity scales as $\csc(\theta/2)$ and is quite large for small twist angle $\theta$. In effect, the $\mathbf{k}$ space periodicity scales as $\theta/2$ and that is reflected in the form of the $\mathbf{q}$ vectors above. We can compare \eqref{blg_ham} with the generic Hamiltonian \eqref{repack}, identifying the operator $D$ of \eqref{Dop} with $\mathfrak{h}$. In continuation of our previous discussion, the operator $D$ is a matrix-valued one, and contains differential operator. Hence if it has to have at least one flat band, it should be supplemented with appropriate periodic boundary condition.

\medskip

\begin{figure}[ht]
   \centering 
    \includegraphics[width=15cm]{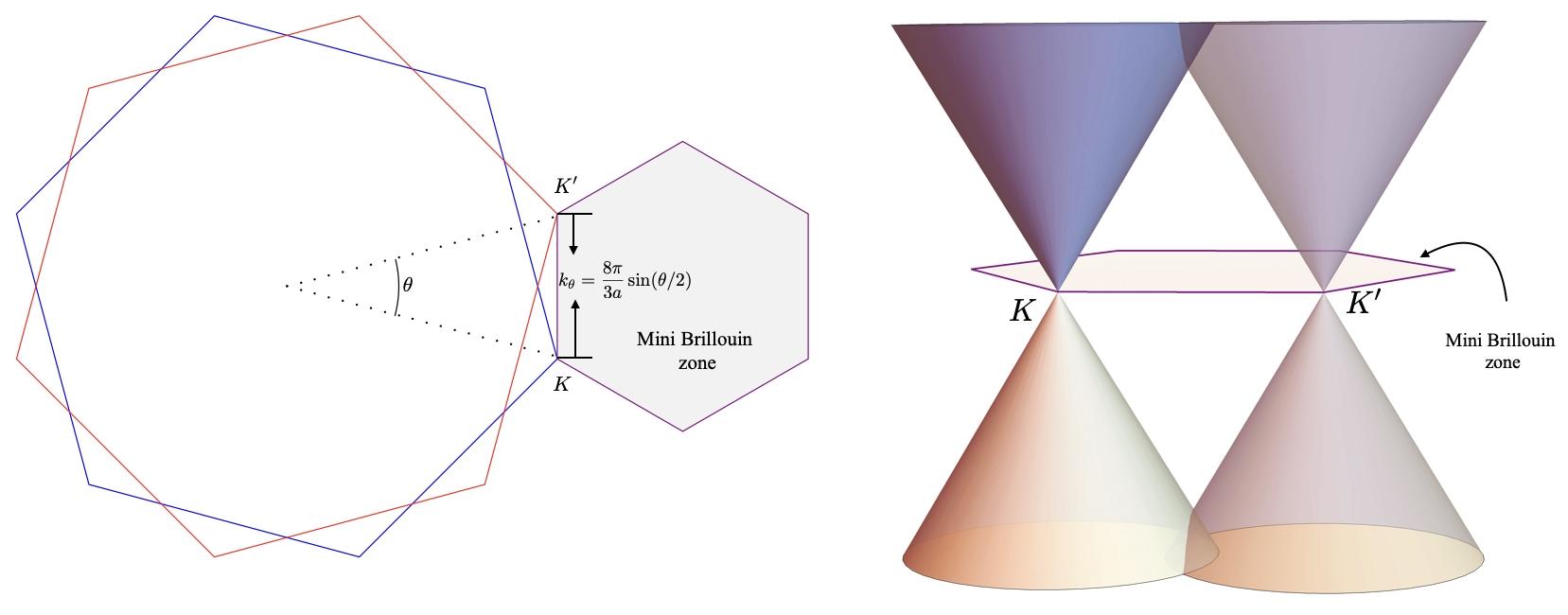} 
    \caption{Couple of Dirac points depicting the dispersion relation in Moire Brillouin zone (momentum space), in absence of inter-layer coupling. The separation between the Dirac points $k_{\theta} = \frac{8 \pi}{3 a} \sin(\theta/2)$ is twist angle dependent. Here $a$ is the lattice parameter. The fermi velocity is, as expected, same as that of monolayer. }
\label{fig1}    
    \end{figure}

Note that by setting $w_1 =0$ in \eqref{blg_ham} one recovers two decoupled Dirac fermions each corresponding to a monolayer graphene sheet. For an angle $\theta$,  in the Fourier space one looks for Bloch waves in smaller `scale', mini-Brillouin zone (mBZ), depicted in the figure \ref{fig1}. In the left panel, the blue and the red hexagons stand for Brillouin zones corresponding to individual mono-layer graphanes, whereas the construction of the mBZ is shown in purple outline. In the absence of inter-layer coupling, the low energy dispersion is given as a pair of filled (sea) and vacant Dirac cones at each vertex (Dirac points) of the mBZ.

We need to keep in mind that introduction of the Moire crystal potential $U(\mathbf{r})$, for arbitrary $\theta$, breaks the space-time symmetries enjoyed previously by the pristine low energy effective theory of monolayer graphene. In the following we will see how those symmetries (and many more) are restored at the magic angle. 

\medskip

For a fixed $\theta$, as one turns on $w_1$ the Dirac cones start to deform. However the gapless nature of the Dirac points (at the points $K, K'$ as per figure \ref{fig1}) is protected by symmetries. This means that one still gets a couple of zero modes corresponding to each of the the points $K$ and $K'$ in Brillouin zone even for nonzero $\alpha$ perturbation in the Hamiltonian \eqref{blg_ham}:
$$ D \Phi_{(K,K')} = 0.$$
In the following, we keep track of the mode $K$, whereas all the properties hold same for the point $K'$ as well. For nonzero $\alpha$ the fermi velocity at the $K$ point is given in terms of the zero-mode:
\begin{eqnarray} \label{ferm_renor}
v_F(\alpha) = \Big |\partial_{\mathbf{k}}\frac{ \Phi^{\dagger}_K G_k  \Phi_K}{\Phi^{\dagger}_K \Phi_K} \Big |_{\mathbf{k} = 0}
\end{eqnarray} 
where $G_k$ is the momentum space Green's function of the operator $D$. One can perturbatively calculate $v_F(\alpha)$ for $\alpha <1$ and up to first order in perturbation, one finds that $v_F ( \alpha = 1/\sqrt3) =0$. This ensures flattening of the erstwhile Dirac cone at the point $ \mathbf{k} =0$, ie the $K$ point. For a value of $w_1$ phenomenologically collected, this happens for the angle $\theta \approx 1.05^{ \circ } $ \cite{tarnopolsky2019origin}, which is famously known as the `magic angle'. 

\medskip

Now we have found 4 zero modes of the operator $D$, to each for the points $K,K'$. If there exists at least one flat-band containing one of those modes, the whole dispersion relation for that band must be identically zero. Hence one next looks for zero modes away from the $K$ point. But the Hamiltonian \eqref{blg_ham} has charge-conjugation (particle-hole) symmetry, which tells us that there will be two such pristine flat bands containing the zero modes at $K$. By virtue of $C_3$ symmetry, the same argument applies to the two other bands containing zero modes of $K'$. Hence if one band flattens, all the 4 of them should follow suit, resulting in vanishing of the Hamiltonian.

\medskip

The essential way to capture this is by looking for zero mode solutions for $D$  throughout the entire mBZ, $i.e.$ finding a basis $ \{\Psi_{\mathbf{k}} | \mathbf{k} \in \mathrm{mBZ}\}$, such that $D \Psi_{\mathbf{k}} (\mathbf{r}) = 0$. The main caveat in finding the kernel of the operator $D$ on the Hilbert space of square integrable 4 component fermions is in boundary conditions. The Moire pattern generated by the twist implies the periodicity: 
\be{}
\mathbf{r} \rightarrow \mathbf{r} + m \mathbf{a}_+ + n\mathbf{a}_- \quad \text{for} \, (m,n) \in \mathbb{Z} \times \mathbb{Z} \quad \text{and} \quad  \mathbf{a}_{\pm} = \dfrac{a}{4\sin(\theta/2)} \left(\pm \sqrt{3},1 \right).
\ee
 This implies that, when brought in the fundamental domain by a $SL(2, \mathbb{Z})$ transformation, the base manifold is a torus of fixed modular parameter $\tau = -1 + \sqrt{3} i$. Also note that the spatial periodicity imposes a discretization in the mBZ points: $\mathbf{k} \cdot \mathbf{a}_{\pm} \in 2 \pi \mathbb{
Z}$.
\medskip 

Now, the Fermi velocity \eqref{ferm_renor} vanishing at some nonzero value of $\alpha$ implies that the kernel of $D$ is spanned by 4 component spinors whose components are given by theta functions with rational characteristic $$\left(\mathbf{k} \cdot \mathbf{a}_{+} /(2 \pi) -1/6, 1/6 -  \mathbf{k} \cdot \mathbf{a}_{-} /(2 \pi)\right).$$ Hence one concludes that for each point on the mBZ, there are non-trivial zero mode of $D$. This implies that the Hamiltonian \eqref{blg_ham} itself vanishes, with all the 4 bands being pristine zero energy ones. In a full fledged numerical calculation, there are other non-flat higher energy bands, not captured by the present model though. However they are substantially gapped from the pristine flat-bands and don't play significant role in determining low energy effective dynamics. Therefore according to our arguments presented in section \ref{connections}, the low energy theory of `magic' angle bilayer graphene is Carrollian invariant.

\medskip

Although it is tempting to conclude that since the Hamiltonian of the system is identically zero, the fermions in question are the massless versions of lower Carrollian fermions, we refrain from making this identification right away as the physics involved in the above is complicated and is crucially dependent on boundary conditions. We have not addressed boundary conditions in our discussions of Carroll fermions and a full blown comparison with magic angle graphene would require us to do so. But the emergent physics is clearly Carrollian and the actors on the stage are fermions, so there is no doubt that a more detailed understanding of Carroll fermions would lead us to a better understanding of magic angle bi-layer graphene. 

%Now we note firstly that the Hamiltonian \eqref{blg_ham} is an integral of a local density $\mathcal{H}(\mathbf{r})$. Secondly, there are no negative modes implying the vanishing of $\mathcal{H}(\mathbf{r})$ at each point of the torus described above and hence $\{\mathcal{H}(\mathbf{r}), \mathcal{H}(\mathbf{r'}) \} = 0$. This conclusively brings us to the point of Carrollian invariance of low energy theory of `magic' angle bilayer graphene.

\subsection*{Unconventional superconductivity}
Observation of emergent superconductivity in magic angle bilayer graphene led experimentalists to test 3,4 and 5 multilayer stacked graphene sheets twisted alternatively at $\pm$ magic angle. Tri-layer \cite{carr2020ultraheavy} onwards \cite{khalaf2019magic}, these systems exhibit flatbands corresponding to Carrollian modes, as well as a set of decoupled Dirac fermions \footnote{Signifying both the signatures of Lorentzian as well as Carrollian physics, decoupled at different energy scales.} and show robust superconductivity \cite{park2021tunable, park2022robust}. This gives rise to an ever stronger conviction regarding the positive role of flatbands  behind robust superconductivity. It is important to note that, although rare in mesoscopic scale physics, flatbands have been observed even in a few systems apart from graphene. What sets graphene apart is the absence of robust emergent superconductivity in these systems, where notably the pristine phenomena of flattening of Dirac cones is absent too.

\medskip

In this era of when a full-blown microscopic theory describing the formation of cooper pairs resulting in this superconductivity is unavailable, we propose from the perspective of global symmetry arguments is that the Carrollian contraction of the Lorentz covariant theory describing low energy of graphene plays a significant role.

\subsection*{Looking ahead}
This is of course just the start and we have identified the symmetry structure associated with generic flat bands to be that of Carroll and in cases of gapless excitations, Conformal Carroll. We remind the reader that these symmetries are infinite dimensional and just as $2d$ relativistic conformal field theories have played a starring role in understanding e.g. the theory of fixed points and phase transitions, it is likely that Carroll and its conformal cousin would be vital to a deeper understanding of the exotic and exciting new physics arising out of flat band structures. 

\newpage

\section{Conclusions and discussions}\label{sec7}

\subsection{Summary of the paper}
In our paper, we have begun an investigation into fermions on Carrollian manifolds and hence, among other things, tried to address what happens if you put fermions on generic null surfaces. 

\medskip

We found that the degenerate nature of Carroll manifolds leads to not one, but two Carrollian Clifford algebras. We investigated the representations of both these versions of Carrollian gamma matrices, constructed explicit matrix representations in $d=4$ and showed how to build up representations of these gamma matrices in arbitrary even and odd dimensions. While the lower gamma matrices furnished a faithful representation of the underlying Carroll algebra, we found that the upper gamma matrices did not. 

\medskip

We called upon charge conjugation to fix this apparent problem and went on to construct actions from both types of gamma matrices. These led us to two distinct theories of Carroll fermions, the lower or time-like fermions and the upper space-like fermions. 

\medskip

The lower fermions were exhibited curious properties: even in higher dimensions, they were effectively one dimensional, they were anti-commuting but did not necessary lead to half integral spins. For massless lower fermions, the Hamiltonian vanished. 

\medskip

The upper fermions were more like usual relativistic fermions. Their action contained both spatial and temporal derivatives. This led to non-vanishing Hamiltonians even in the massless limit. The fermions did not behave lower dimensional and had half integral spins. 

\medskip

In perhaps the most interesting part of our paper, we argued that Carrollian physics naturally emerges in condensed matter systems where one encounters the physics of flat bands, that are in vogue. We argued that the flattening of energy bands essentially means lightcones closing up in position space and the emergence of Carroll. We then explicitly showed Carroll invariance in these theories in general.  

\medskip

We then went on to give explicit examples of Carrollian fermions in condensed matter systems. We showed that a particular fermionic lattice model in 1+1 dimensions exactly reduces to our lower fermions. We also showed how Carrollian physics would show up in magic bi-layer graphene. 

\subsection{Future directions}
There are numerous directions that this work opens up. Let us list a few of them below. 

\begin{itemize}

\item{\em{Covariantisation:}} We have looked at fermionic theories on flat Carroll backgrounds. One of the immediate things to do is to find a covariant formulation in terms of the underlying geometric data and see how fermions couple of non-flat Carroll backgrounds. 

\item{\em{Understanding the limit:}} As we briefly discussed earlier, it is of importance to understand the Carrollian limit of relativistic fermions and figure out how the structures we discussed in this paper arises naturally from their relativistic parent theories. This would give more insight to the origin of the two different species of fermions we have found in the Carrollian world. Clarification as to whether the lower and upper fermions are electric and magnetic theories in the current nomenclature would also help clarify some properties of the theories we have discovered from the intrinsic analysis. 

\item{\em{Dimensional reduction:}} We discussed that the lower fermions behaved like one-dimensional objects. It would be useful to quantify this further. Investigation of the spin-statistics theorem for Carrollian theories may hold important clues in this regard. It is possible that the lower fermions are like anyons in the relativistic set up. Relatedly, it would be of interest to investigate whether there is a reduction of our upper fermions. Recently it has been shown that the spacelike mass Carroll scalar theory can be reduced to a relativistic Euclidean CFT in one lower dimension \cite{Baiguera:2022lsw}. Whether a similar thing holds for the spacelike or upper fermions would be rather intriguing to work out. 

\item{\em{More condensed matter applications:}} We have just begun to uncover Carroll symmetry in condensed matter systems. Even the case of twisted bi-layer graphene which we discussed in this paper needs to be understood better by looking at non-trivial boundary conditions. It would be exciting to look at the physics of e.g. quantum Hall systems. Again the possible connection to lower fermions which behave somewhat like anyons is tantalizing. 

\item{\em{Supersymmetry:}} Supersymmetry in Carrollian systems, especially in conformal Carrollian ones, have been studied from the point of view of the algebra and different contractions of the parent relativistic algebras. See e.g. \cite{Bagchi:2022owq} for a recent account of this in generic dimensions and references therein. But now, armed with the knowledge of the Carroll Clifford algebra, and the properties of Carroll fermions, it will be instructive to revisit this and construct supersymmetry in the Carroll case from scratch. This is work in progress. 

\end{itemize} 

Clearly there are numerous very interesting things to follow up on. Of these, the emerging connections to real life condensed matter examples are most intriguing.

\newpage

\subsection*{Acknowledgements}
It is a pleasure to thank Kinjal Banerjee, Sudipta Dutta, Daniel Grumiller, Marc Henneaux, Blagoje Oblak, Marios Petropoulos for helpful conversations. Some preliminary results were presented at the 2nd Carroll Workshop at the University of Mons, Belgium and AB thanks the participants for discussions. 

\medskip

The work of AB was partially supported by a Swarnajayanti fellowship (SB/SJF/2019-20/08) from the Science and Engineering Research Board (SERB) India, the SERB grant (CRG/2020/002035), a Research-in-Groups grant at the Erwin Schr{\"o}dinger Institute, Vienna and a visiting professorship at Ecole Polytechnique Paris. AB also acknowledges the warm hospitality of Vienna University of Technology and the Niels Bohr Institute, Copenhagen during various stages of this work. The work of ArB is supported by the Quantum Gravity Unit of the Okinawa Institute of Science and Technology Graduate University (OIST). ArB would like to thank TU Wien for hospitality during the early stages of this project. The work of RB is supported by the following grants from the SERB, India: CRG/2020/002035, SRG/2020/001037. SM is supported by grant number 09/092(1039)/2019-EMR-I from Council of Scientific and Industrial Research (CSIR).

	\newpage
	\appendix
	\section*{APPENDICES} 
	\section{Carroll Isometries}\label{CI}
	For Carroll case the vector fields $X$ obey 
	\begin{eqnarray}
		\mathcal{L}_{\xi}\tilde{h}_{ab}=0,\quad \mathcal{L}_{\xi}\theta_a=0
	\end{eqnarray}
	Now from the definition of lie derivative
	\begin{equation*}
		\mathcal{L}_{\xi}\tilde{h}_{ab} = \xi^c\p_c \tilde{h}_{ab} + \tilde{h}_{cb}\p_a \xi^c + \tilde{h}_{ac}\p_b \xi^c
	\end{equation*}
	Using the expression of $\tilde{h}_{ab}$ we get, 
	\begin{eqnarray}
		\p_A \xi_B + \p_B \xi_A &=& 0 \qquad {A=1,2,...,d} \\
		\p_A \xi^B + \p_B \xi^A &=& 0 
	\end{eqnarray}
	In the last line we have used $\xi_A = \delta_{AB}\xi^B$. Also from
	\begin{equation}
		\mathcal{L}_{\xi}\theta^a = \xi^b\p_b\theta^a - \theta^b\p_b \xi^a
	\end{equation}
	using $\theta^a = -\delta_a^0$, we get
	\begin{eqnarray}
		\p_0 \xi^0 =0, \quad \p_0 \xi^A =0.
	\end{eqnarray}
	Solving these equations give 
	\be{}
	\xi^0 = f(x^k) + a, \quad \xi^A = \omega^A_Bx^B + b^A.
	\ee 
	So the isometry group of the degenerate Carrollian metric is infinite-dimensional.
	If one restricts the transformation to be linear, one gets the finite-dimensional Carroll group.
%	From the third condition
%	\begin{equation}
%		\mathcal{L}_{\xi}\Gamma = \p_a\p_b \xi^c = 0
%	\end{equation}
%	$\xi^0$ is linear in $x$ and $\xi^A$ has to be linear in $x$.
	So the vector field for the finite dimensional Carroll group
	\begin{eqnarray}
		\xi=\xi^0\p_0 + \xi^A\p_A = a\p_t + \omega^A_Bx^B\p_A + v_Ax^A\p_t + b^A\p_A
	\end{eqnarray}
	For Conformal case the equations are 
	\begin{eqnarray}
		\mathcal{L}_{\xi}\tilde{h}=\lambda \tilde{h},\quad \mathcal{L}_{\xi}\theta= -\frac{\lambda}{2}\theta
	\end{eqnarray}
	By solving in the similar manner we get 
	\begin{eqnarray}
		\p_A \xi_B + \p_B \xi_A &=& \lambda\delta_{AB} \qquad {A=1,2,...,d} \\
		\p_0 \xi^0 &=& -\frac{\lambda}{2}, \quad \p_0 \xi^A = 0
	\end{eqnarray}
One can find the solution as 
	\begin{eqnarray}
		\xi= f(x)\p_t + \omega^A_Bx^B\p_A + b^A\p_A + \Delta(t\p_t + x^A\p_A) +\nonumber\\ \kappa_A\big(2x^A(t\p_t + x^A\p_A)-x^C x_C\delta^{AB}\p_B\big)
\end{eqnarray}		
Here $b$,$\Delta$,$\kappa$ are arbitrary coefficients , $\omega^A_B$ is antisymmetric.

	\newpage

\newpage
\bibliographystyle{JHEP}
\bibliography{NLF}	
	
\end{document}